\PassOptionsToPackage{unicode}{hyperref}
\PassOptionsToPackage{hyphens}{url}
\PassOptionsToPackage{dvipsnames,svgnames,x11names}{xcolor}
\PassOptionsToPackage{format=plain,
  labelfont={bf,sf,small,singlespacing},
  textfont={sf,small,singlespacing},
  justification=justified,
  margin=0.25in}{caption}%
\documentclass[
  11pt,
]{article}
\usepackage{xcolor}
\usepackage[lmargin=1in,rmargin=1in,tmargin=1in,bmargin=1in]{geometry}
\usepackage{amsmath,amssymb}
\usepackage{iftex}
\ifPDFTeX
  \usepackage[T1]{fontenc}
  \usepackage[utf8]{inputenc}
  \usepackage{textcomp} 
    \usepackage[p,osf,swashQ,scale=1.02]{cochineal}
  \usepackage{amsthm}
  \usepackage[cochineal,vvarbb]{newtxmath}
      \usepackage[scaled]{helvet}
    \usepackage[scale=0.95,varl]{inconsolata}
\else 
    \usepackage{unicode-math} 
    \usepackage{mathpazo}
  \usepackage{newpxtext}
    \usepackage[scaled]{helvet}
    \defaultfontfeatures{Scale=MatchLowercase}
  \defaultfontfeatures[\rmfamily]{Ligatures=TeX,Scale=1}
\fi

\usepackage{sectsty}
\ifPDFTeX\else
\fi
\IfFileExists{upquote.sty}{\usepackage{upquote}}{}
\IfFileExists{microtype.sty}{
  \usepackage[]{microtype}
  \UseMicrotypeSet[protrusion]{basicmath} 
}{}
\makeatletter
\@ifundefined{KOMAClassName}{
  \IfFileExists{parskip.sty}{%
    \usepackage{parskip}
  }{
    \setlength{\parindent}{0pt}
    \setlength{\parskip}{6pt plus 2pt minus 1pt}}
}{
  \KOMAoptions{parskip=half}}
\makeatother

\usepackage{longtable,booktabs,array}
\usepackage{calc} 
\usepackage{etoolbox}
\makeatletter
\patchcmd\longtable{\par}{\if@noskipsec\mbox{}\fi\par}{}{}
\makeatother
\IfFileExists{footnotehyper.sty}{\usepackage{footnotehyper}}{\usepackage{footnote}}
\makesavenoteenv{longtable}
\usepackage{graphicx}
\makeatletter
\newsavebox\pandoc@box
\newcommand*\pandocbounded[1]{
  \sbox\pandoc@box{#1}%
  \Gscale@div\@tempa{\textheight}{\dimexpr\ht\pandoc@box+\dp\pandoc@box\relax}%
  \Gscale@div\@tempb{\linewidth}{\wd\pandoc@box}%
  \ifdim\@tempb\p@<\@tempa\p@\let\@tempa\@tempb\fi
  \ifdim\@tempa\p@<\p@\scalebox{\@tempa}{\usebox\pandoc@box}%
  \else\usebox{\pandoc@box}%
  \fi%
}
\def\fps@figure{htbp}
\makeatother

\providecommand{\tightlist}{%
  \setlength{\itemsep}{0pt}\setlength{\parskip}{0pt}}

\usepackage[]{natbib}
\usepackage{mathtools}
\usepackage{amssymb}
\usepackage[italicdiff]{physics}

\newcommand{\F}{\mathcal{F}}
\newcommand{\G}{\mathcal{G}}
\AtBeginDocument{\renewcommand{\H}{\mathcal{H}}}
\newcommand{\X}{\mathcal{X}}
\newcommand{\loocv}{\mathrm{LOO}}

\renewcommand{\op}{\mathrm{op}}

\let\span\undefined

\DeclareMathOperator{\span}{span}
\newcommand{\supp}{\mathrm{supp}}

\newcommand{\XZ}{\underline{\vb{XZ}}}
\newcommand{\XZj}{\widetilde{\vb{XZ}}_j}
\newcommand{\XZjm}{\widetilde{\vb{XZ}}_{mj}}
\newcommand{\XZjmp}{\widetilde{\vb{XZ}}'_{mj}}
\newcommand{\xz}{\underline{\vb{xz}}}
\newcommand{\xzt}{\widetilde{\vb{xz}}}

\newcommand{\R}{\ensuremath{\mathbb{R}}}

\newcommand{\half}{\frac{1}{2}}
\newcommand{\thalf}{\tfrac{1}{2}}

\newcommand{\tquart}{\tfrac{1}{4}}

\renewcommand{\bar}{\overline}
\renewcommand{\hat}{\widehat}
\newcommand{\eps}{\varepsilon}
\newcommand{\iid}{\overset{\text{iid}}{\sim}}

\DeclareMathOperator{\E}{\mathbb{E}}
\renewcommand{\Pr}{\mathbb{P}}
\DeclareMathOperator{\Var}{\mathbb{V}}
\newcommand{\indep}{\mathbin{\perp\!\!\!\!\!\:\perp}}
\newcommand{\ind}{\mathbf{1}}
\newcommand{\cvp}{\to^p}
\newcommand{\cvd}{\Rightarrow}

\DeclareMathOperator{\Norm}{\mathcal{N}}

\usepackage{threeparttable}

\makeatletter
\newcommand{\assumplabel}[2]{%
   \protected@write \@auxout {}{\string\newlabel{#1}{{#2}{\thepage}{#2}{#1}{}}}%
   \hypertarget{#1}{#2}%
}
\makeatother
\newenvironment{assump}[2][]{%
\par\medskip\noindent%
\def\tempa{}\def\tempb{#1}%
\lowercase{\def\lblkey{asm-#2}}%
\textbf{Assumption \assumplabel{\lblkey}{#2}}%
\ifx\tempa\tempb\else~(#1)\fi%
\textbf{.}\itshape%
}{\medskip\normalfont}
\makeatletter
\@ifpackageloaded{caption}{}{\usepackage{caption}}
\AtBeginDocument{%
\ifdefined\contentsname
  \renewcommand*\contentsname{Table of contents}
\else
  \newcommand\contentsname{Table of contents}
\fi
\ifdefined\listfigurename
  \renewcommand*\listfigurename{List of Figures}
\else
  \newcommand\listfigurename{List of Figures}
\fi
\ifdefined\listtablename
  \renewcommand*\listtablename{List of Tables}
\else
  \newcommand\listtablename{List of Tables}
\fi
\ifdefined\figurename
  \renewcommand*\figurename{Figure}
\else
  \newcommand\figurename{Figure}
\fi
\ifdefined\tablename
  \renewcommand*\tablename{Table}
\else
  \newcommand\tablename{Table}
\fi
}
\@ifpackageloaded{float}{}{\usepackage{float}}
\floatstyle{ruled}
\@ifundefined{c@chapter}{\newfloat{codelisting}{h}{lop}}{\newfloat{codelisting}{h}{lop}[chapter]}
\floatname{codelisting}{Listing}

\usepackage{amsthm}
\theoremstyle{plain}
\newtheorem{lemma}{Lemma}[section]
\theoremstyle{plain}
\newtheorem{corollary}{Corollary}[section]
\theoremstyle{plain}
\newtheorem{proposition}{Proposition}[section]
\theoremstyle{plain}
\newtheorem{theorem}{Theorem}[section]
\theoremstyle{remark}
\AtBeginDocument{}

\makeatother
\makeatletter
\@ifpackageloaded{caption}{}{\usepackage{caption}}
\@ifpackageloaded{subcaption}{}{\usepackage{subcaption}}
\makeatother
\usepackage{bookmark}
\IfFileExists{xurl.sty}{\usepackage{xurl}}{} 
\hypersetup{
  pdftitle={Estimation of Conditional Means from Aggregate Data},
  pdfauthor={Cory McCartan; Shiro Kuriwaki},
  pdfkeywords={aggregate data, ecological inference, double/debiased
machine learning},
  colorlinks=true,
  linkcolor={black},
  filecolor={Maroon},
  citecolor={VioletRed4},
  urlcolor={DodgerBlue4},
  pdfcreator={LaTeX via pandoc}}

\makeatletter
\newcommand\@shorttitle{}
\newcommand\shorttitle[1]{\renewcommand\@shorttitle{#1}}
\usepackage{fancyhdr}
\fancyheadoffset{0pt}
\makeatother
\renewenvironment{abstract}{
  \centerline
  {\large\sffamily\bfseries Abstract}\vspace{-0.25em}
  \begin{quote}\small
}{
  \end{quote}
}

\title{\sffamily\bfseries\huge\parfillskip=0pt
\rightskip=0pt plus .5\textwidth
\leftskip=0pt plus .5\textwidth
\emergencystretch=.3\textwidth Identification and Semiparametric
Estimation of Conditional Means from Aggregate Data}
\shorttitle{Estimation of Conditional Means from Aggregate Data}
\author{\textbf{Cory McCartan}\footnote{
To whom correspondence should be addressed.
Email: \texttt{\href{mailto:mccartan@psu.edu}{mccartan@psu.edu}}.
Website: \url{https://corymccartan.com}.
Address:
325 Thomas Building, 461 Pollock Rd, University Park, PA 16802.
The authors thank Gary King, Kosuke Imai, James Bailie, and Benjamin K.
Bechtold for helpful comments and discussion.}
\\Department of Statistics%
\\Pennsylvania State University%
\vspace{2pt}
 \and \textbf{Shiro Kuriwaki}
\\Department of Political Science%
\\Yale University%
\vspace{2pt}
 }
\date{April 30, 2026}

\begin{document}
\allsectionsfont{\sffamily}

\maketitle

\begin{abstract}
We introduce a new method for estimating the mean of an outcome variable
within groups when researchers only observe the average of the outcome
and group indicators across a set of aggregation units, such as
geographical areas. Existing methods for this problem, also known as
ecological inference, implicitly make strong assumptions about the
aggregation process. We first formalize weaker conditions for
identification which hold conditionally on covariates. To efficiently
control for many covariates, we propose a debiased machine learning
estimator that is based on nuisance functions restricted to a partially
linear form. Our estimator admits a semiparametric sensitivity analysis
which allows researchers to evaluate the impact of violations of the key
identifying assumption. We also propose a nonparametric test for the
identifying assumption itself. Finally, we derive asymptotically valid
confidence intervals for local, unit-level estimates under additional
assumptions. Simulations and validation on real-world data where ground
truth is available demonstrate the advantages of our approach over
existing methods. Open-source software is available which implements the
proposed methods.
\end{abstract}

\textbf{\textit{Keywords}}\quad aggregate data~\textbullet~ecological
inference~\textbullet~double/debiased machine learning

\section{Introduction}\label{sec-intro}

One of the most common statistical tasks is estimating the mean of an
outcome \(Y\) within subgroups defined by a discrete variable \(X\). In
many settings, however, researchers do not jointly observe \(Y\) and
\(X\) for each unit, but only observe the average of each variable
within a grouping variable such as geography. Despite the simplicity of
the aggregation operation, the problem of estimating the conditional
mean \(\E[Y\mid X]\) from marginal means \(\bar Y\) and \(\bar X\),
known as \emph{ecological inference}, is far from straightforward.

For example, consider estimating how different racial and income groups
(\(X\)) are exposed to pollution (\(Y\)), while only observing the
fraction of individuals in each racial or income group in a ZIP code
(\(\bar X\)) and the average pollution exposure in the ZIP code
(\(\bar Y\)) \citep{jbaily2022air}. Because joint information about
\(X\) and \(Y\) within ZIP codes is not observed, the conditional mean
\(\E[Y\mid X]\) cannot be point-identified from the data without further
assumptions \citep{cho2008cross}. This is a common challenge in
epidemiology, where exposure and disease are measured separately
\citep{greenland1994invited}. Another well-studied instance arises in
voting rights litigation, which requires estimating the voting behavior
within racial subgroups, while only observing precinct-level election
returns and Census statistic on race \citep{greiner2006ecological}.
Finally, statistical agencies generally report demographic, economic,
and public health data only in aggregate form, due to data collection or
privacy constraints. A long line of work has therefore tackled the
ecological inference problem, beginning with
\citet{robinson1950ecological} and Goodman
\citetext{\citeyear{goodman1953ecological}; \citeyear{goodman1959some}}.

This paper develops a new understanding of and methodology for the
ecological inference problem. We propose a semiparametrically efficient
estimator that allows researchers to minimize bias by controlling for
many covariates. Our approach formalizes previously implicit
assumptions, and improves on the issues of confounding and computational
efficiency in past work.

Existing methods largely fall into two categories. One, applicable only
when \(Y\) is bounded, focuses on partial identification of
\(\E[Y\mid X]\) by deriving bounds on the estimand under various
assumptions
\citep{duncan1953alternative, cross2002regressions, fan2016estimation, manski2018credible, jiang2020ecological}.
In practice these bounds are often too wide; intervals for different
levels of \(X\) almost always overlap in practice.\footnote{ Some
  authors
  \citep[e.g.,][]{judge20047, muzellec2017tsallis, bontemps2025functional}
  have proposed selecting a point from the partial identification region
  according to an ad-hoc criterion, such as entropy minimization or
  divergences based on optimal transport. While providing a single
  estimate, these proposals lack statistical or substantive
  justification, and as such, it is not possible to quantify their bias
  or uncertainty.}

The other approach, following Goodman, aims for point identification
under often-unstated assumptions, generally relying on parametric
regression models. The seminal work of \citet{king1997solution} combined
the regression framework with the partial identification bounds in a
Bayesian varying coefficient model, spurring numerous extensions
\citep{rosen2001bayesian, wakefield2004ecological, james2009r}, though
the approach was somewhat controversial \citep{freedman1998solution}.

Our proposed method addresses two general challenges in this existing
literature. First, the necessary identifying assumptions are rarely
stated explicitly and are often extremely strong
\citep{freedman1998solution}; in practice, users of ecological inference
methods rarely evaluate the plausibility of their
assumptions.\footnote{ An exception to the pattern is
  \citet{imai2008bayesian}, who presented two possible parametric
  identifying assumptions. Nevertheless, awareness among practitioners
  of the necessary identifying assumptions remains low.} Second, most
methods rely on strong parametric assumptions, are computationally
intensive, and lack any inferential guarantees.\footnote{ The
  computational limitations may have discouraged practitioners from
  controlling for essential covariates. Existing methods have
  complicated likelihood functions that require computationally
  intensive inference methods such as Markov chain Monte Carlo (MCMC)
  algorithms, and exhibit significant slowdowns as the number of
  covariates increases. \citet{km2025review} discuss these computational
  challenges in more detail.}

To address the identification ambiguity, we formalize the ecological
inference problem in Section~\ref{sec-ident} with an explicit model for
the aggregation process, which allows us to relate individual-level and
aggregate-level data. Under this framework, we state two main
identifying assumptions, one at the individual level and one at the
aggregate level, and prove they are sufficient for identification of
\(\E[Y\mid X]\) in aggregate data. These identification results
highlight the importance of aggregate-level covariates \(Z\) in making
the identifying assumptions more plausible. These results also make
clear the role that the number of individuals in each aggregation unit
plays in identification and estimation, an aspect of the problem that
prior literature has largely ignored.

To address the restrictive parametric specifications, we propose in
Section~\ref{sec-est} a new double/debiased machine learning estimator
for \(\E[Y\mid X]\) that allows researchers to minimize bias by
controlling for many covariates. Compared to existing estimation
approaches, our proposed semiparametric estimator is statistically and
computationally efficient without making strong parametric assumptions,
and achieves good accuracy and coverage in practice, as we demonstrate
in simulations and validation on real-world data where ground truth is
available (Section~\ref{sec-valid}).

Key to the estimator's development is a result on the conditional
expectation function (CEF) \(\E[\bar Y\mid \bar X, Z]\) of the aggregate
outcome: under the identification assumptions, we show that the CEF
takes a partially linear form. This connection also enables application
of the Riesz representation theorem, which highlights the importance of
an additional \emph{positivity} assumption for ecological inference that
is rarely recognized in the literature. Analogously to causal inference,
positivity essentially requires sufficient variation in \(\bar X\) after
controlling for covariates.

We also introduce three new tools that make ecological inference more
useful and more reliable (Section~\ref{sec-extend}). First, in addition
to our estimation theory for the so-called \emph{global} estimand
\(\E[Y\mid X]\), we develop asymptotically valid confidence intervals
for the \emph{local} estimands \(\E[Y\mid X, G=g]\), the conditional
means within each aggregation unit \(g\). While not point-identified,
these local estimands are often of interest to practitioners. Second, we
introduce both a sensitivity analysis and a hypothesis test for the key
identifying assumption. These are particularly valuable for applied
researchers, who often are concerened about the plausibility of the
identifying assumption in practical settings. We adapt a sensitivity
analysis framework in causal inference \citep{chernozhukov2022sens} to
provide the first sensitivity analysis for aggregate data. Third, we
provide in the appendix a test for the identifying assumption that draws
on the testable implication of a partially linear CEF. This test, which
has no analogue in the missing data context, is nevertheless approximate
and has limited power, and so we recommend use of the sensitivy analysis
primarily.

Our proposed semiparametric estimator and these three tools are all
implemented in open-source software \citep{seine}, which we apply in a
demonstration in Section~\ref{sec-appl} to the air pollution data of
\citet{jbaily2022air}. Together with the formalization of identifying
assumptions, these methods place ecological inference on a more
practical and robust foundation.

\section{Identification}\label{sec-ident}

Consider a population of exchangeable individuals \(i=1,\dots,n\), each
belonging to an aggregation unit \(G_i\in\G\), which we will refer to as
\emph{geographies} herein for simplicity, since in most applications the
aggregation units correspond to geographic areas. We write the
population of each geography as \(N_g\coloneq |\{i:G_i=g\}|\). Each
individual has a continuous outcome variable \(Y_i\in\R\) and a
categorical predictor variable \(X_i\in\{0, 1\}^d\) with
\(d\coloneq |\X|\) levels, represented as a vector of \(d\) mutually
exclusive indicator variables for each possible level in \(\X\). For
cases where \(Y\) is discrete, one can apply the methods here to the
indicator variable for each level of \(Y\) separately. We assume
throughout that \(\E[Y_i^2]<\infty\).

\subsection{Aggregation model and
estimand}\label{aggregation-model-and-estimand}

Rather than observing \(X_i\) and \(Y_i\) for each individual, the
researcher observes the aggregated variables \[
    \bar{Y}_g \coloneq \frac{1}{N_g}\sum_{i\,:\,G_i=g} Y_i \qand
    \bar{X}_g \coloneq \frac{1}{N_g}\sum_{i\,:\,G_i=g} X_i.
\]

One of the challenges in studying inference is the need to work with
both individual-level and aggregate-level data simultaneously;
observations that are i.i.d. at one level are not i.i.d. at the other
level, in general. To aid in working across levels, we introduce random
indices over individuals and geographies, which will allow us to
compactly write aggregations and regressions as expectations over these
random indices. For \(\omega\in\R^n\) an arbitrary vector of individual
weights with \(\E[\omega_i]=1\), define a random index \(I^\omega\) by
\(\Pr(I^\omega=i)=\omega_i / n\). Then we can define
\(G^\omega\coloneq G_{I^\omega}\) to be a random index over geographies.
We will primarily work with two special cases. First, when
\(\omega_i\propto N_{G_i}^{-1}\), so that \(G^\omega\) is uniform over
the geographies, we will drop the superscript and simply write \(G\).
Second, when all \(\omega_i=1\), we will use \(I^n\) and \(G^n\), so
\(I^n\) is uniform over the \(n\) individuals, and \(\Pr(G^n=g)\) is
proportional to \(N_g\).

In this notation, we may write \(\bar Y_g=\E_n[Y_I\mid G=g]\)
\footnote{We could have equivalently used \(I_n\) here, since both \(I\)
  and \(I^n\) are uniform conditional on geography.} and the global mean
\(\bar Y=\E_n[Y_{I^n}]=\E_n[{\bar Y}_{G^n}]\), where \(\E_n\) denotes an
expectation over the empirical measure. In addition to being more
compact, the random index notation will permit us to state the main
identification results without making an i.i.d. or superpopulation
assumption about the aggregate-level data.\footnote{ The estimation
  results require an asymptotic framework, for which we adopt an i.i.d.
  model of geographies.}

Associated with each observed \((\bar X_g, \bar Y_g)\) is a vector of
unobserved regression coefficients \[
B_g \coloneq \E_n[X_I X_I^\top\mid G=g]^{-1}\E_n[X_I Y_I\mid G=g],
\] which represent the (sample) mean value of \(Y\) for each group in
\(\X\). By definition, \(\bar Y_g\), \(\bar X_g\), and \(B_g\) are
connected by the law of total expectation, traditionally referred to in
ecological inference as the \emph{accounting identity}:
\begin{equation}\protect\phantomsection\label{eq-acct-id}{
    \bar Y_g = B_g^\top\bar X_g.
}\end{equation}

Finally, there may be covariates \(Z_g\) available at the geography
level. Together, \((Z_g, \bar X_g, B_g)\) are the full data at the
aggregate level; the researcher observes only the coarsened
\((Z_g, \bar X_g, \bar Y_g)\).

The \emph{global estimand} is the vector of individual-level conditional
means \(\beta\),\footnote{Often, the sample equivalent of this estimand,
  where \(\E\) is replaced by \(\E_n\), is of interest. The
  identification arguments go through identically, but proving
  estimation results requires a superpopulation and asymptotic
  framework.} defined by \[
\begin{aligned}
    \beta_j \coloneq&
    \E[Y_{I^n} \mid X_{I^nj}=1]
    = \frac{\E[N_{G_I} Y_I \mid X_{I^nj}=1]}{\E[N_{G_I}\mid X_{I^nj}=1]} \\
    =& \frac{\E[\bar X_{G^nj} B_{G^nj}]}{\E[\bar X_{G^nj}]}
    = \frac{\E[N_G \bar X_{Gj} B_{Gj}]}{\E[N_G \bar X_{Gj}]}.
\end{aligned}
\] Here, we have written both representations---as a conditional average
of the individual \(Y_i\), and as a weighted average over the
\(B_g\)---in terms of both the individual-weighted random indices
\(G^n\) and the geography-weighted random indices \(G\). We next
investigate under what conditions \(\beta\) is identified from the
coarsened data \((Z_g, \bar X_g, \bar Y_g)\).

\subsection{Identification at the aggregate and individual
level}\label{identification-at-the-aggregate-and-individual-level}

Eq.~\ref{eq-acct-id} makes clear the fundamental identification
challenge: each observation \((\bar X_g, \bar Y_g)\) brings with it
\(d\) unknown parameters: the entries of \(B_g\). This makes
Eq.~\ref{eq-acct-id} a type of random-coefficient model, albeit one with
no error term. These models are well-studied
\citep[e.g.,][]{beran1992estimating}, and to identify \(\beta\), some
kind of regularity across the \(B_g\) must be assumed. For example, if
there is no variation in \(B_g\), so that each \(B_g=\beta\), then there
is a single \(d\)-dimensional unknown parameter, which can be estimated
via linear regression. In fact, assuming constancy across \(B_g\) is
stronger than necessary. What is required is that variation in \(B_g\)
be unrelated to variation in \(\bar X_g\); constancy is a special case
of this condition. The following assumption formalizes the condition;
while \citet{beran1992estimating} state the assumption without
covariates, it can be easily weakened to hold conditional on covariates.

\begin{assump}[Coarsening at random, uniform over individuals]{CAR-U}
For all \(\bar x\) and \(z\),
\(\E[B_{G^n}\mid Z_{G^n}=z, \bar X_{G^n}=\bar x]=\E[B_{G^n}\mid Z_{G^n}=z]\).\end{assump}

This assumption is a familiar analogue of the ignorability assumption in
causal inference or the missing-at-random assumption in missing data
analysis. As in \citet{heitjan1991ignorability}, coarsening at random
(CAR) means that the variable which determines the amount of coarsening
or information loss, \(\bar X_{G^n}\), is (mean) independent of the
unobserved data \(B_{G^n}\), given covariates. Since \(B_{G^n}\) is not
observed, in general it is not possible to directly check whether
Assumption~\ref{asm-car-u} holds in the data at hand: researchers should
rely on their substantive knowledge. However, as we will see,
Assumption~\ref{asm-car-u} implies a certain modeling restriction which
may be testable from data. We propose a test for the assumption and
discuss its limitations in Appendix~\ref{sec-id-test}.

Because of the weighting by \(N_g\), Assumption~\ref{asm-car-u} is best
interpreted at the individual level: that for an individual \(i\)
selected uniformly at random, knowing the average \(\bar X_{G_i}\) in
their geography \(G_i\) does not change the expectation of the
individual's corresponding \(B_{G_i}\), given the covariates
\(Z_{G_i}\). Since the assumption is an individual-level one stated in
terms of aggregate variables, it may be difficult to interpret. The
following weighted version of the assumption yields a more helpful
interpretation.

\begin{assump}[Coarsening at random]{CAR}
For all \(\bar x\), \(k\), and \(z\),
\(\E[B_G\mid Z_G=z, \bar X_G=\bar x, N_G=k]=\E[B_G\mid Z_G=z]\).\end{assump}

This assumption can of course be stated in two stages: first, that
\(\bar X\) is mean-independent of \(B\) given \(Z\) and \(N\), and
second, that \(B\) is mean-independent of \(N\) given \(\bar X\) and
\(Z\). Although Assumption~\ref{asm-car} is slightly stronger than
Assumption~\ref{asm-car-u}, we will use it throughout the rest of the
paper due to its easier interpretability and the flexibility it provides
in estimation. If only Assumption~\ref{asm-car-u} but not
Assumption~\ref{asm-car} holds, then estimation can proceed identically,
but with observations weighted by \(N_g\) throughout.

Previous work which used an aggregate-level setup often took
\(\bar X_g\) and \(N_g\) as fixed, and so did not consider the ways in
which \(N_g\) could be correlated with other variables. For example,
\citet{ansolabehere1995bias} claimed that weighting by \(N_g\) was
necessary for unbiased estimation, but note that in practice weighting
did not seem to make a large difference. This is the case because
weighting is only required when \(N_g\) is related to \(B_g\) even after
controlling for \(\bar X_g\) and \(Z_g\). As the next result shows, in
general, either Assumption~\ref{asm-car-u} or Assumption~\ref{asm-car}
is sufficient for identification of \(\beta\). All proofs are deferred
to Appendix~\ref{sec-app-proofs}.

\begin{theorem}[Nonparametric
identification]\protect\hypertarget{thm-id}{}\label{thm-id}

For all \(j\in\X\), under Assumption~\ref{asm-car-u}, \[
\beta_j = \frac{\E[\bar X_{G^nj}\E[\bar Y_{G^n}\mid Z_{G^n}, \bar X_{G^nj}=1]]}{\E[\bar X_{G^nj}]},
\] and under Assumption~\ref{asm-car}, \[
\beta_j = \frac{\E[N_G\bar X_{Gj}\E[\bar Y_G\mid Z_G, \bar X_{Gj}=1]]}{\E[N_G\bar X_{Gj}]},
\]

\end{theorem}

Note that when there are no covariates, i.e., \(Z\) is null,
Assumption~\ref{asm-car} is strong and implausible. In the air pollution
example, Assumption~\ref{asm-car} without covariates would imply that a
low-income resident of Los Angeles and a low-income resident of rural
Montana would have the same average \(\text{PM}_{2.5}\) exposure. In a
setting where \(Y\) is vote choice and \(X\) is race,
Assumption~\ref{asm-car} without covariates would imply that white
voters' preferences are identical between Seattle, Wash. and a heavily
Republican rural county such as Fairmount, Ga. Thus, in most
applications, it will be critical to include relevant covariates that
explain variation in the \(B_g\), so that Assumption~\ref{asm-car} is
more plausible. Where covariates are available at the individual level,
they can be aggregated, either marginally or jointly, to form \(Z_g\).
For example, in the voting setting, individual-level age and sex may be
available from the voter file, and their contingency table at the
precinct level could be included as a covariate.

One additional difficulty in evaluating the plausibility of
Assumption~\ref{asm-car} is that it is stated in terms of the aggregated
data itself, while the estimand itself is defined at the individual
level. In some contexts, it may be more straightforward to make
identifying assumptions at the individual level. As
Theorem~\ref{thm-car-ind-agg} records, the following assumption is
sufficient for Assumption~\ref{asm-car}.

\begin{assump}[Coarsening at random at the individual level]{CAR-IND}
For every individual \(i\), and for each \(g\), \(x\), and \(z\),
\(\E[Y_i\mid G_i=g, X_i=x, Z_{G_i}=z]=\E[Y_i\mid X_i=x, Z_{G_i}=z]\).\end{assump}

\begin{theorem}[Identification at individual
level]\protect\hypertarget{thm-car-ind-agg}{}\label{thm-car-ind-agg}

Assumption~\ref{asm-car-ind} \(\implies\) Assumption~\ref{asm-car}.

\end{theorem}

Because \(G\) appears directly in Assumption~\ref{asm-car-ind}, it may
be particularly helpful when researchers have substantive knowledge of
the process that assigns individuals to aggregation units (geographies).
However, it is a stronger assumption than Assumption~\ref{asm-car}.
There may be situations where Assumption~\ref{asm-car-ind} does not hold
while Assumption~\ref{asm-car} does.

\section{Estimation}\label{sec-est}

In this section, we apply Assumption~\ref{asm-car} and
Theorem~\ref{thm-id} to develop a semiparametrically efficient estimator
for \(\beta\). We begin with an observation about the form of the
conditional expectation function (CEF) \(\gamma_0\) of \(\bar Y\) under
Assumption~\ref{asm-car}:
\begin{equation}\protect\phantomsection\label{eq-cef}{
\gamma_0(\bar X, Z) := \E[\bar Y_G\mid Z_G, \bar X_G]
= \E[B_G^\top\bar X_G\mid Z_G, \bar X_G]
= \eta_0(Z_G)^\top \bar X_G,
}\end{equation} where \(\eta_0(Z_G):=\E[B_G\mid Z_G]\). Thus, without
any parametric assumptions, \(\gamma_0\) belongs to a restricted class
of partially linear functions \[
\Gamma := \{(\bar x, z) \mapsto \eta(z)^\top \bar x: \{\eta_j\}_{j\in\X}\in L^2(Z)\}.
\] Clearly, \(\Gamma\) is a linear subspace of \(L^2(Z_G, \bar X_G)\);
below, we will show that under an additional assumption, \(\Gamma\) is
in fact a \emph{closed} linear subspace. First, however, we discuss
estimation when Assumption~\ref{asm-car} holds without covariates.

\subsection{Estimation without
covariates}\label{estimation-without-covariates}

The first ecological inference methods were based on simple linear
regression of \(\bar Y_G\) on \(\bar X_G\)
\citep{goodman1953ecological, goodman1959some}. When \(Z\) is null, we
can express \(Y_G\) as \[
\bar Y_G =\eta^\top \bar X_G + \eps_G^\top\bar X_G,
\] where \(\eps_G=B_G-\E[B_G\mid Z_G]\) is the projection residual from
Eq.~\ref{eq-cef}. Because
\(\E[B_G\mid Z_G]=\E[B_G\mid Z_G, \bar X_G, N_G]\), \(\eps_G\) is
orthogonal to any function of \((Z_G, \bar X_G, N_G)\), and we have
immediately that \[
\E[\eps_G^\top\bar X_G \mid\bar X_G]
=\E[\eps_G^\top \mid\bar X_G]\bar X_G=0.
\] Thus \(\eta\) can be estimated efficiently by least squares, and
since it is constant, \(\beta=\eta\). When only
Assumption~\ref{asm-car-u} holds, the least-squares regression must be
weighted by \(N_G\), optionally multiplied by a function of
\(\bar X_G\), to guarantee unbiasedness; when Assumption~\ref{asm-car}
holds, any weights which are a function of \(N_G\) and \(\bar X_G\) can
be used. Slightly weaker conditions for finite-sample unbiasedness of
least squares in this setting are possible; see
\citet{ansolabehere1995bias} for an analysis when \(d=2\).

When \(Y\) is binary, so \(\bar Y\) is bounded, a least squares
estimator does not incorporate information contained in these bounds,
which \citet{duncan1953alternative} and \citet{king1997solution} argue
can be substantial. When \(d=2\), \citet{king1997solution} explicitly
models the random coefficients \(B_G\) in order to incorporate the
bounds on \(\bar Y_G\). Specifically, he takes
\(B_G=\eta+\eps_G\sim \Norm_{[0,1]^2}(\mu,\Sigma)\), where the subscript
indicates truncation to the unit square. This ensures that
\(0\le \bar Y_G\le 1\), and allows for Bayesian inference for each
\(B_G\). However, it does impose a strong parametric assumption on the
distribution of \(\eps_G\).

Thus it is clear that both Goodman's regression and King's method are
fully consistent with the accounting identity Eq.~\ref{eq-acct-id}, and
they both implicitly assume Assumption~\ref{asm-car} holds
unconditionally. The key difference is in the treatment of the error
term \(\eps_G\): Goodman's regression is semiparametric, in that it is
agnostic to the distribution of the error term; King, by contrast, makes
a distributional assumption. However, this assumption provides several
benefits: while Goodman regression only estimates
\(\eps_G^\top\bar X_G\), King's estimates \(\eps_G\) directly and
ensures it respects any bounds on \(Y\). This allows for estimates of
each geography's \(B_G\) which are consistent with the accounting
identity Eq.~\ref{eq-acct-id} and may be partially identified due to
bounds on \(Y\).

King's model runs into computational difficulties when \(d>2\), since
the normalizing constant and moments of a truncated Normal distribution
are not easily available in higher dimensions \citep[see][ for
details]{km2025review}. Moreover, while King allows for the linear
inclusion of a covariate \(Z\), he does not discuss the challenges of
modeling \(Z\) flexibly, as is required to avoid misspecification bias.

\subsection{Semiparametric estimation with
covariates}\label{sec-est-semi}

By Theorem~\ref{thm-id}, we can write the global estimand using the
notation in Eq.~\ref{eq-cef} as
\begin{equation}\protect\phantomsection\label{eq-id-functional}{
\beta_j = \E[\gamma_0(e_j, Z_G)u(N_G, \bar X_{Gj})],
}\end{equation} where \(e_j\) is a standard basis vector and
\(u(N_G, \bar X_{Gj})=N_GX_{Gj}/\E[N_GX_{Gj}]\) weights by the size of
group \(j\) in each geography. Our overall estimation strategy,
following \citet{chernozhukov2022riesz}, is to rewrite \(\beta_j\) in
Neyman-orthogonal form using a Riesz representation of \(\beta_j\),
estimate nuisance functions flexibly using a series estimator, and then
combine the nuisance estimates to form a semiparametrically efficient
estimator for \(\beta_j\).

Consider the Eq.~\ref{eq-id-functional} as a mapping \(\Gamma\to\R\). We
can easily see that \(\gamma\mapsto \gamma(e_j, Z)u(N, \bar X_j)\) is
linear in \(\gamma\), so \(\beta_j\) is a linear functional of
\(\gamma\). To apply the Riesz representation theorem to this
functional, we require two additional assumptions.

\begin{assump}[Positivity]{POS}
The random variables \((\bar X_G, Z_G)\) have joint density
\(f(\bar x, z)\) with respect to a dominating measure on
\(\Delta^d\times\supp(Z_G)\), where \(\Delta^d\) is the
\(d\)-dimensional simplex, and there exists a \(\delta>0\) such that
\(f(\bar x, z)>\delta f(z)\) for all
\((\bar x, z)\in\Delta^d\times\supp(Z_G)\), where \(f(z)\) is the
marginal density of \(Z_G\).\end{assump}

\begin{assump}[Bounded N]{BND}
There exists a \(C_N<\infty\) with \(N_GX_{Gj}\le C_N\E[N_GX_{Gj}]\) for
each \(j\).\end{assump}

Assumption~\ref{asm-bnd} bounds \(u\), ensuring that no single geography
dominates the estimand. Assumption~\ref{asm-pos}, while more involved to
state, essentially requires that there be sufficient variation in
\(\bar X_G\) after controlling for \(Z_G\). It is analogous to the
positivity or overlap assumption in causal inference. Note that the
existence of a joint density \(f(x, z)\) and its positivity at the
vertices of the simplex is also sufficient for the uniqueness of the
conditional expectations \(\gamma(e_j, Z)\), which are evaluated at a
measure-zero set; Assumption~\ref{asm-pos} is in practical terms,
therefore, a mild strengthening of this requirement.

These two assumptions establish two key results: first, that the
partially-linear function class \(\Gamma\) is a \emph{closed} linear
subspace of \(L^2(\bar X_G, Z_G)\); and second, that \(\beta_j\) is
mean-square continuous in \(\gamma\). In what follows, we write
\(\norm{\cdot}\) for the \(L^2(\bar X_G, Z_G)\) norm. These results
yield an immediate corollary, due to the Riesz representation theorem.

\begin{proposition}[]\protect\hypertarget{prp-subspace}{}\label{prp-subspace}

Under Assumption~\ref{asm-pos}, \(\Gamma\) is a closed linear subspace
of \(L^2(\bar X_G, Z_G)\).

\end{proposition}

\begin{proposition}[]\protect\hypertarget{prp-msc}{}\label{prp-msc}

Under Assumption~\ref{asm-pos} and Assumption~\ref{asm-bnd}, for each
\(x\in\X\) the mapping
\(\gamma\mapsto\E[\gamma(e_j, Z_G)u(N_G, \bar X_{Gj})]\) is mean-square
continuous in \(\gamma\), i.e.,
\(\E[\gamma(e_j, Z_G)u(N_G, \bar X_{Gj})]\le C_{\Pr}\norm{\gamma}^2\)
for a \(C_{\Pr}<\infty\) depending on \(\Pr\).

\end{proposition}

\begin{corollary}[]\protect\hypertarget{cor-rr}{}\label{cor-rr}

For each \(j\in\X\), there exists a unique
\(\alpha_{0j}(\bar X_G, Z_G)=\zeta_{0j}(Z_G)^\top\bar X_G \in\Gamma\)
with \(\norm{\alpha_{0j}}^2<\infty\) and satisfying
\(\beta_j=\E[\alpha_{0j}\gamma_0]=\E[\alpha_{0j} \bar Y]\).

\end{corollary}

We refer to \(\alpha_{0j}\) as the \emph{Riesz representer} of
\(\beta_j\); critically, it also belongs to the restricted class
\(\Gamma\). While it is defined implicitly, in
Appendix~\ref{sec-app-riesz} we present a closed-form expression for
\(\alpha_{0j}\) as a weighted log-derivative of the conditional density
\(f(\bar x\mid z)\). We further discuss its interpretation in the
context of our sensitivity analysis in Section~\ref{sec-sens}.

The second moment of the Riesz representer is tied directly to the
modulus of continuity that establishes Proposition~\ref{prp-msc}, which
in turn hinges critically on Assumption~\ref{asm-pos}. The larger the
second moment of \(\alpha_{0j}\), the less variation in \(\bar X_{gj}\)
there is conditional on \(Z_g\), and the greater the risk of
Assumption~\ref{asm-pos} not holding. This is analogous to causal
inference, where the distribution of the propensity scores plays a
similar role in assessing overlap.

Corollary~\ref{cor-rr} implies that we could estimate \(\beta_j\) in two
ways: either by estimating \(\gamma_0\) and then plugging into
Eq.~\ref{eq-id-functional}, or by estimating \(\alpha_{0j}\) and
plugging into \(\E[\alpha_{0j} \bar Y]\). However, since both
\(\gamma_0\) and \(\alpha_{0j}\) are functions which must be estimated,
either approach can lead to significant regularization biases. Instead,
a Neyman-orthogonal representation of \(\beta_j\) can be formed based on
the efficient influence function of \(\beta_j\)
\citep{newey1994asymptotic}:
\begin{equation}\protect\phantomsection\label{eq-neyman}{
\beta_j(\gamma_0,\alpha_{0j}) = \E[\gamma_0(e_j, Z_g)u(N_G,\bar X_G) +
    \alpha_{0j}(\bar X_G, Z_G)(\bar Y_G - \gamma_0(\bar X_G, Z_G))].
}\end{equation} This representation is robust to small errors in either
nuisance function, in the sense that its Gateaux derivative with respect
to the nuisance functions vanishes, i.e.,
\(\partial_\gamma \beta_j(\gamma_0,\alpha_{0j}) = \partial_{\alpha_j} \beta_j(\gamma_0,\alpha_0) = 0\)
\citep{chernozhukov2022riesz}. This Neyman orthogonality property is
closely related to double robustness: in fact, if \(\gamma\) is properly
specified in Eq.~\ref{eq-neyman}, then even with a misspecified
\(\alpha\), the score in Eq.~\ref{eq-neyman} is still unbiased for
\(\beta_j\), and vice versa. This can be easily seen by applying
iterated expectations and the representing property of \(\alpha_{0j}\).

Let \(\hat{\gamma}_m\) and \(\hat{\alpha}_{mj}\) be estimates of
\(\gamma_0\) and \(\alpha_{0j}\) based on \(m:=|\G|\) geographies. Then
the proposed estimator for \(\beta_j\) is
\begin{equation}\protect\phantomsection\label{eq-beta-hat}{
\begin{aligned}
\hat\beta_{mj} = \beta_{mj}(\hat\gamma_m, \hat\alpha_{mj})
&:= \frac{1}{m}\sum_{g\in\G} \psi_{gj}(\bar Y_g, N_g, \bar X_g, Z_g, \hat\gamma_m, \hat\alpha_{mj});\\
\psi_{gj}(\bar Y_g, N_g, \bar X_g, Z_g, \hat\gamma_m, \hat\alpha_{mj})
    &:= \hat\gamma_m(e_j, Z_g)u(N_g,\bar X_g) +
    \hat\alpha_{mj}(\bar X_g, Z_g)(\bar Y_g - \hat\gamma_m(\bar X_g, Z_g)).
\end{aligned}
}\end{equation}

In the next subsections, we discuss estimation of the nuisance functions
\(\gamma_0\) and \(\alpha_{0j}\) and the statistical properties of
\(\hat\beta_{mj}\). As we noted in Section~\ref{sec-ident}, to state the
asymptotic results for the estimation of the nuisance function and for
\(\hat\beta_{mj}\), we need an asymptotic framework that treats the
geographies as i.i.d. Consequently, in the remainder of the paper we
index the geographies by \(g\) rather than \(G\) and rely on the
following assumption.

\begin{assump}[]{IID}
\((\bar Y_g, \bar X_g, Z_g, N_g)\iid\Pr\).\end{assump}

This differs from the random-index setup of Section~\ref{sec-ident},
where the individuals are i.i.d. but the geographies are in general not.
We stress that the assumption is introduced only to allow a meaningful
asymptotic analysis of the estimation procedure; it is not required for
identification. Rather than interpreting the i.i.d. assumption
literally, one should think of the remaining results as meaningfully
characterizing the behavior of our estimators insofar as the set of
observed geographies can be treated \emph{as if} i.i.d. Our validation
studies demonstrate that this is a reasonable assumption in practice.

\subsection{\texorpdfstring{Estimation of \(\gamma_0\) and
\(\alpha_0\)}{Estimation of \textbackslash gamma\_0 and \textbackslash alpha\_0}}\label{sec-nuisance}

As discussed in \citet{chernozhukov2022riesz}, \(\alpha_{0j}\) can be
estimated via the following representation:
\begin{equation}\protect\phantomsection\label{eq-min-alpha}{
\begin{aligned}
\alpha_{0j} &= \arg\min_{\alpha\in\Gamma} \E[(\alpha - \alpha_{0j})^2]
= \arg\min_{\alpha\in\Gamma} \E[\alpha^2 - 2\alpha_{0j}\alpha + \alpha_{0j}^2] \\
&= \arg\min_{\alpha\in\Gamma} \E[\alpha^2(\bar X_g, Z_g) - 2\alpha(e_j, Z_g)u(N_g, \bar X_{Gj})].
\end{aligned}
}\end{equation} where the final step follows by the representation
property of \(\alpha_{0j}\) and since \(\alpha_{0j}\) is fixed.

In principle, \(\gamma_0\) and \(\alpha_{0j}\) could be therefore
estimated using any nonparametric regression or machine learning method,
with \(\gamma_0\) estimated by minimizing a squared-error loss, and
\(\alpha_{0j}\) estimated by minimizing the loss in
Eq.~\ref{eq-min-alpha}. However, these methods would not produce
estimates \(\hat\gamma_m\) and \(\hat\alpha_{mj}\) which belong to the
restricted class \(\Gamma\). To fully leverage the restriction to
\(\Gamma\), we propose series estimators that are in \(\Gamma\) by
construction.

A \emph{linear sieve basis} is a sequence \(\{\Phi_m\}_{m=1}^\infty\) of
vectors of uniformly bounded functions
\(\Phi_m=(\phi_{mk}\in L^\infty(Z_g))_{k=1}^{J_m}\) of dimension
\(J_m\). We discuss three possible sieve bases in
Appendix~\ref{sec-bases}, including interactions of polynomials in each
variable, tensor-product splines, and a more recent tensor-product
cosine basis \citep{zhang2023regression}, as well as the number of basis
functions needed to attain the rates discussed here.

By interacting the elements of a particular sieve basis with the
components of \(\bar X\), we form a basis for a subspace of \(\Gamma\):
\[
\Gamma_m := \{(\bar x, z)\mapsto (\bar x\otimes \Phi_m(z))^\top\theta : \theta\in\R^{dJ_m}\},
\] where \(\otimes\) is the Kronecker product, i.e., all pairwise
interactions between the elements of \(\Phi_m\) and \(\bar x\). Our
proposed estimators for \(\gamma_0\) and \(\alpha_{0j}\) are then the
series ridge regression estimators
\begin{equation}\protect\phantomsection\label{eq-sieve-est}{
\begin{aligned}
\hat\gamma_m(\lambda) &:= \arg\min_{\theta}\{ \E_m[(\bar Y_g -
    (\bar X_g\otimes\Phi_m(Z_g))^\top\theta)^2] + \lambda \theta^\top\theta \} \qand \\
\hat\alpha_{mj}(\lambda) &:= \arg\min_{\theta}\{ \E_m[
    ((\bar X_g\otimes\Phi_m(Z_g))^\top\theta)^2
    - 2u(N_g, \bar X_{Gj})(e_j\otimes \Phi_m(Z_g))^\top\theta
] + \lambda \theta^\top\theta \},
\end{aligned}
}\end{equation} where \(\E_m\) denotes the empirical expectation. It is
clear that
\(\hat\gamma_m(\lambda), \hat\alpha_{mj}(\lambda)\in\Gamma_m\subseteq\Gamma\).
The closed-form solution for \(\hat\gamma_m(\lambda)\) is well-known; we
present a closed-form solution for \(\hat\alpha_{mj}(\lambda)\) in
Appendix~\ref{sec-app-riesz}. To estimate \(\lambda\), we employ
leave-one-out cross-validation (LOOCV) for the loss of \(\hat\gamma_m\).
The LOOCV loss can be efficiently computed for a range of \(\lambda\)
using the hat matrix and the singular value decomposition of the design
matrix. The asymptotic considerations in \citet{singh2024kernel} suggest
that the same \(\lambda\) can be used for both \(\hat\gamma_m\) and
\(\hat\alpha_{mj}\); we do so here, since it is not computationally
convenient to compute LOOCV errors for \(\hat\alpha_{mj}\). We discuss
in Appendix~\ref{sec-app-impl} several other practical considerations in
implementing the estimator: bounded \(Y\), the use of weights in
estimation, and the choice of sieve basis.

\subsubsection{Convergence rates}\label{convergence-rates}

For the estimate \(\hat\beta_{mj}\) to be asymptotically normal, we will
need \(\hat\gamma_m\) and \(\hat\alpha_{mj}\) to converge quickly enough
to their targets, which requires conditions on the true \(\eta_0(z)\) as
well as on the sieve basis \(\Phi_m\). Sieve estimators for varying
coefficient models have been proposed and analyzed before
\citep{park2015varying}. However, most treatments focus on regression
estimators and do not directly apply to estimating \(\alpha_{0j}\).
Moreover, Assumption~\ref{asm-pos} and Assumption~\ref{asm-bnd} can be
sufficient for various regularity conditions required by other
estimators, and the fact that \(\bar X\) is supported on the simplex
\(\Delta^d\) is specific to this case as well. Thus we state and prove
the necessary conditions and results in full here.

\begin{assump}[Sieve estimation regularity conditions]{SR}
We assume the following about the data-generating process:

\begin{enumerate}
\def\labelenumi{(\arabic{enumi})}
\tightlist
\item
  \(Z_g\) is supported on a bounded subset of \(\R^p\) for some \(p\),
  and has a bounded density \(0<f(z)<\infty\) with respect to Lebesgue
  measure on its support.
\item
  \(\Var[Y_g\mid Z_g=z, \bar X_g=\bar x]\) is bounded on
  \(\Delta^d\times\supp(Z_g)\).
\item
  There exists a function class \(\F\subseteq L^2(Z_g)\) with
  \(\eta_0,\zeta_{0j}\in\F^d\) and \(\norm{\eta_{0j}}_\infty<\infty\)
  and \(\norm{\zeta_{0jk}}_\infty<\infty\) for each \(j,k\in\X\), where
  \(\eta_0\) and \(\zeta_{0j}\) are the true component functions for
  \(\gamma_0\) and \(\alpha_{0j}\), respectively.
\end{enumerate}

For the conditions on the sieve basis \(\Phi_m\), let \(\nu\) denote
Lebesgue measure on \(\supp(Z_g)\), rescaled to have unit mass, and let
\[
\rho_{m} := \sup_{f'\in\F}\inf_{f\in\span\Phi_m} \norm{f - f'}_{2,\nu} \qand
A_m := \sup_{f\in\span\Phi_m, \norm{f}_{2,\nu}\neq 0} \norm{f}_\infty / \norm{f}_{2,\nu},
\] where \(\norm{f}_{2,\nu}^2 := \int f^2 d\nu\). We assume

\begin{enumerate}
\def\labelenumi{(\arabic{enumi})}
\setcounter{enumi}{3}
\tightlist
\item
  \(\span\Phi_m\) is identifiable, i.e.,
  \(\norm{f}_{2,\nu}=0 \implies f=0\) for \(f\in\span\Phi_m\);
\item
  \(A_m\rho_{m}\to 0\); and
\item
  \(A_m^2J_m/m\to 0\).
\end{enumerate}\end{assump}

Condition (1) ensures that the \(L^2(\bar X_G,Z_G)\) norm is equivalent
to the Lebesgue norm \(\norm{\cdot}_{2,\nu}\) on \(\supp(Z_G)\), which
means that properties of the sieve basis can be checked on the latter,
independent of the data distribution, and carry over to the former.
Conditions (4)--(6) hold for many common sieve bases and function
classes, as we discuss below. These conditions would generally permit
estimation of functions in \(\F\) at the rate \(\rho_m + \sqrt{J_m/m}\);
we, however, need to estimate functions in the space
\(\Gamma^\F := \{(\bar x, z)\mapsto f(z)^\top\bar x : f\in\F^d\}\subseteq \Gamma\)
which contains \(\gamma_0\) and \(\alpha_{0j}\) under
Assumption~\ref{asm-sr} (3). The next theorem shows that this is
possible at the same rate.

\begin{theorem}[]\protect\hypertarget{thm-sieve}{}\label{thm-sieve}

Under Assumption~\ref{asm-bnd}, Assumption~\ref{asm-pos}, and
Assumption~\ref{asm-sr}, \(\hat\gamma_m\) and \(\hat\alpha_{mj}\) exist
uniquely with probability approaching one as \(m\to\infty\), and for all
\(j\in\X\), we have for the unpenalized estimators that \[
\norm{\hat\gamma_m(0) - \gamma_0} = O_{\Pr}(\rho_m + \sqrt{J_m/m}) \qand
\norm{\hat\alpha_{mj}(0) - \alpha_{0j}} = O_{\Pr}(\rho_m + \sqrt{J_m/m}).
\] If additionally, the eigenvalues of the matrix
\(\Phi_m(\vb Z)^\top\Phi_m(\vb Z)\) (where \(\vb Z\) is the matrix of
\(Z_g\)) are uniformly bounded away from zero, and
\(\lambda_m=O(\sqrt{J_m/m})\), then \[
\norm{\hat\gamma_m(\lambda_m) - \gamma_0} = O_{\Pr}(\rho_m + \sqrt{J_m/m}) \qand
\norm{\hat\alpha_{mj}(\lambda_m) - \alpha_{0j}} = O_{\Pr}(\rho_m + \sqrt{J_m/m}).
\]

\end{theorem}

The rate on \(\lambda_m\) ensures that the penalty does not bias the
estimator asymptotically. In practice, as mentioned above, we pick the
penalty using LOOCV. When the design matrix satisfies certain
conditions, such as being a linear transformation of i.i.d. variables,
\citet{patil2021uniform} show that LOOCV is uniformly consistent for the
optimal penalty, even when the number of predictors grows with the
sample size. While their setup differs from the one here, LOOCV is very
likely to reduce the estimation error in finite samples from the
unpenalized estimator, and so \(\hat\gamma_m(\hat\lambda_\loocv)\) and
\(\hat\alpha_{mj}(\hat\lambda_\loocv)\) should converge at the same
rate.

\subsection{Properties of the proposed estimator}\label{sec-conv}

The proposed estimator \(\hat\beta_{mj}\) in Eq.~\ref{eq-beta-hat} is a
double/debiased machine learning (DML) estimator
\citep{chernozhukov2022riesz}, but unlike many DML estimators, the
nuisance functions \(\gamma_0\) and \(\alpha_{0j}\) are estimated on the
same data as \(\hat\beta_{mj}\) rather than being cross-fitted. While
cross-fitting makes theoretical analysis easier, in finite samples cross
fitting over just \(k\) folds can substantially increase variance. Some
authors recommend generating multiple sets of \(k\) folds, which reduces
variance but increases computational cost. Here, \(\gamma_0\) and
\(\alpha_{0j}\) are estimated using a ridge penalty, and so we might
expect better behavior than an estimator using a black-box machine
learning method that could be arbitrarily sensitive to the data used to
fit it. We therefore establish the asymptotic normality and
semiparametric efficiency of \(\hat\beta_{mj}\) without cross-fitting,
using the results of \citet{chen2022debiased}, which rely on an
algorithmic stability condition that is satisfied by the ridge penalty
used here.

\begin{theorem}[]\protect\hypertarget{thm-dml}{}\label{thm-dml}

Suppose that \(\hat\gamma_m\) and \(\hat\alpha_{mj}\) are the
ridge-penalized series estimators in Eq.~\ref{eq-sieve-est} that achieve
estimation error rates
\(\norm{\hat\gamma_m - \gamma_0} = o_{\Pr}(m^{-1/4})\) and
\(\norm{\hat\alpha_{mj} - \alpha_{0j}} = o_{\Pr}(m^{-1/4})\) for each
\(j\in\X\). Assume also that \(\lambda\asymp \sqrt{J_m/m}\), that the
eigenvalues of the matrix \(\Phi_m(\vb Z)^\top\Phi_m(\vb Z)\) are
uniformly bounded away from zero, and that \(\norm{\bar Y}_{2r}<\infty\)
for some \(r>1\). Then \(\hat\beta_m\) is asymptotically normal with
limiting distribution \[
\sqrt{m}(\hat\beta_m - \beta) \cvd \Norm(0, \E[\psi_g\psi_g^\top]),
\] where \(\psi_g\) is the vector of scores \(\psi_{gj}\) for each
\(j\in\X\).

\end{theorem}

The semiparametric efficiency of \(\hat\beta_m\) follows because
\(\psi_g\) is the efficient influence function for \(\beta\) and
\(\E[\psi_g\psi_g^\top]\) is the semiparametric efficiency bound.

Theorem~\ref{thm-dml} means that in practice we can easily construct
asymptotically valid confidence regions for \(\beta\) using the sample
covariance of the scores \(\psi_g\). This also allows for asymptotically
valid confidence intervals for linear contrasts of \(\beta\), which are
often of interest in applications measuring differences in \(Y\) between
groups.

\section{Further Extensions}\label{sec-extend}

In this section, we discuss two extensions of the proposed method:
estimation of the \emph{local} quantities \(B_g\) for each geography
\(g\), and a sensitivity analysis for violations of
Assumption~\ref{asm-car}. A third extensions, a possible hypothesis test
for Assumption~\ref{asm-car}, is presented in
Appendix~\ref{sec-id-test}.

\subsection{Local estimates}\label{sec-local}

Often, researchers are interested not just in the global estimand
\(\beta=\E[Y\mid X]\), but how this relationship varies by geography.
For example, in political science, \(\beta\) may describe the voting
preference (\(Y\)) of a racial group (\(X\)) nationally, but researchers
may also be interested in how this relationship varies across counties
or precincts. These local quantities are exactly the missing data
\(B_G\). Since there is a single (unobserved) \(B_G\) per geography, it
is of course not possible to consistently estimate the \(B_G\)
themselves. However, it is possible to construct valid confidence
regions \(B_G\), under additional assumptions.

Under Assumption~\ref{asm-car}, \(B_G := \eta_0(Z_G) + \eps_G\), with
\(\eps_G\) mean-zero and mean-independent of \((N_G, \bar X_G, Z_G)\).
Thus, a natural point estimate for \(B_G\) is \(\hat\eta(Z_G)\). The
more variation in \(B_G\) (and thus \(\bar Y_G\)) explained by \(Z_G\),
the more accurate this point estimate will be. However, it will not be
consistent for \(\beta_G\) since \(\eps_G\) has non-zero variance.
Additionally, while \(B_G\) must satisfy the accounting identity
(Eq.~\ref{eq-acct-id}), i.e., \(\bar Y_G=B_G^\top \bar X_G\), in general
the estimates \(\hat\eta(Z_G)\) will not. We aim to develop a point
estimate and confidence region for \(B_G\) that addresses these two
issues. Doing so will require consistently estimating the covariance
matrix \(\Var[\eps_G]\), which requires additional assumptions.

\begin{assump}[Coarsening at random, second moments]{CAR2}
For all \(\bar x\), \(k\), and \(z\),
\(\E[B_GB_G^\top\mid X_G=\bar x, Z_G=\bar z, N_G=k]=\E[B_GB_G^\top\mid Z_G=z]\).\end{assump}

Assumption~\ref{asm-car2} could be equivalently written in terms of the
residuals \(\eps_G\). Of course, both Assumption~\ref{asm-car2} and
Assumption~\ref{asm-car} are implied by the stronger condition that
\(B_G\) is conditionally independent of \(N_G\) and \(\bar X_G\) given
\(Z_G\). A version of Assumption~\ref{asm-car2} that does not condition
on \(N_G\) could also be applied, analogously to
Assumption~\ref{asm-car-u}.

Denote the covariance matrix as \(\Sigma\) and let
\(\Sigma(z):=\Var[\eps_G\mid Z_G=z]\). Under Assumption~\ref{asm-car2},
\(\Sigma(z)\) fully describes the conditional variance structure of
\(\eps_G\). Additionally, let
\(\kappa_0(x,z) :=\E[(\bar Y_G - \E[\bar Y_G \mid \bar X_G,Z_G])^2\mid \bar X_G=x,Z_G=z]\).\\
We then have the following identification result.

\begin{proposition}[]\protect\hypertarget{prp-var-id}{}\label{prp-var-id}

Under Assumption~\ref{asm-car2}, for any \(j,k\in\X\) we have
\(\Sigma_{jk}(z) = 2(\kappa_0(\thalf e_j + \thalf e_k, z) - \tquart\kappa_0(e_j, z) - \tquart\kappa_0(e_k, z))\).

\end{proposition}

The form of this result is due to the polarization identity for
recovering a bilinear form from a quadratic form. The result in
Proposition~\ref{prp-var-id} means that a consistent estimate of
\(\kappa_0\) can be used along with a consistent estimate of
\(\gamma_0\) (which includes \(\eta_0\)) to form an asymptotically valid
confidence region for \(B_G\) using the multivariate Chebyshev
inequality. As discussed above, however, the point estimate
\(\hat\eta(Z_G)\) will not in general satisfy the accounting identity.

To further improve the point estimate and confidence region, we can
project the estimate onto the \(d-1\)-dimensional region implied by the
accounting identity. Specifically, let
\(H(\bar x, \bar y) := \{b\in \R^d : b^\top \bar x = \bar y\}\) be the
set of possible values of \(B_G\) given \(\bar X_G=\bar x\) and
\(\bar Y_G=\bar y\) for an unbounded \(\bar Y_G\); we let \([H_g]\)
denote the matrix with columns forming an orthonormal basis
\(H(\bar X_g, \bar Y_g)\). Such a basis can be efficiently computed by
taking the QR decomposition of the block matrix \((\bar X_g\quad I_d)\)
and discarding the first column of \(Q\). Also let \(\hat\Sigma(z)\) be
the estimate of \(\Sigma(z)\) obtained from the expression in
Proposition~\ref{prp-var-id} using an estimated
\(\hat\kappa\).\footnote{ In finite samples, the \(\hat\Sigma(z)\)
  yielded by Proposition~\ref{prp-var-id} may not be positive
  semidefinite. In these cases, we recommend projecting
  \(\hat\Sigma(z)\) onto the space of positive semidefinite matrices,
  e.g., by setting all negative eigenvalues to zero. This will not
  affect the consistency of \(\hat\Sigma(z)\).} Then we can obliquely
project \(\hat\eta(Z_g)\) onto \(H(\bar X_g, \bar Y_g)\) along
\(\hat\Sigma(Z_g)\) to obtain a point estimate \(\tilde B_g\) that
satisfies the accounting identity: \[
\hat B_g := \hat\Pi_g \hat\eta(Z_g); \quad
\hat\Pi_g := [H_g]([H_g]^\top \hat\Sigma(Z_g)^{-1}[H_g])^{-1}[H_g]^\top \hat\Sigma(Z_g)^{-1}.
\] where \(\hat\Pi_g\) is the oblique projection matrix. We can then
further obliquely project \(\hat B_g\) onto
\(\supp(B_G)=\supp(\bar Y_G)^d\), which is convex, yielding a point
estimate \(\hat B'_g\) that lies in \[
H'(\bar x, \bar y) := H(\bar x, \bar y)\cap \supp(\bar Y_G)^d.
\] Of course, when \(Y\) is unbounded, \(H'=H\). Now define a confidence
region \[
\begin{aligned}
{R'_g}^\alpha &:= \{b\in H'(\bar X_g,\bar Y_g) :
    (b - \hat B'_g)^\top (\hat\Pi_g\hat\Sigma(Z_g))^+ (b - \hat B'_g) \le \frac{d-1}{\alpha} \},
\end{aligned}
\] where \(A^+\) denotes the Moore-Penrose pseudoinverse of \(A\). Then
we have the following result.

\begin{theorem}[]\protect\hypertarget{thm-local-ci}{}\label{thm-local-ci}

Suppose that \(\hat\eta\cvp\eta_0\) and \(\hat\kappa\cvp\kappa_0\)
pointwise. Then for \(0<\alpha<1\), as \(m\to\infty\),
\(\Pr(B_g\in {R'_g}^\alpha) \ge 1-\alpha + o(1)\).

\end{theorem}

In practice, we might expect these confidence regions to be
conservative, especially for bounded \(Y\) when a second projection is
used. Additional distributional assumptions on \(\eps_G\), such as
unimodality or Normality, can be used to further tighten the confidence
regions in practice. For example, for confidence intervals for a single
component \(B_{Gj}\), if the distribution of
\(\eps_{G}^\top\bar X_G\mid Z_G\) is assumed unimodal, then the width of
the confidence interval can be reduced by a factor of \(2/3\)
\citep{vysochanskij1980justification}.

\citet{breunig2021varying} discusses estimation of other aspects of the
distribution of \(\eps\) in varying coefficient models, such as higher
moments or quantiles, in more detail, and develops sieve estimators for
efficiently estimating these quantities. These estimators may be applied
to build intervals for \(B_G\), which in some cases may be narrower, but
more work is needed to apply these in a way that respects the accounting
identity, as the intervals developed in this section do.

\subsection{Sensitivity analysis}\label{sec-sens}

Every result so far has relied critically on the
Assumption~\ref{asm-car} assumption. In practice, it is unlikely that
Assumption~\ref{asm-car} holds \emph{exactly}, and so it is important to
understand how sensitive the proposed estimates of \(\beta\) are to
violations of this assumption. To do so, we can apply results of
\citet{chernozhukov2022sens}, who develop a nonparametric sensitivity
analysis for estimands for which a Riesz representer exists.

Rather than assume that Assumption~\ref{asm-car} holds, the sensitivity
analysis assumes that it holds conditional on an unobserved variable
\(A_G\). This is not really an additional assumption, since we can
always take \(A_G=\E[B_G\mid N_G, \bar X_G, Z_G]\), which then makes
Assumption~\ref{asm-car} conditional on \(A_G\) hold trivially.

Let \(\gamma^A_0\) and \(\alpha^A_{0j}\) be the regression function and
Riesz representer, respectively, defined conditional on \(A_G\). Unlike
\(\gamma_0\) and \(\alpha_{0j}\), which can be estimated from the data,
\(\gamma^A_0\) and \(\alpha^A_{0j}\) cannot be estimated because \(A_G\)
is not observed. However, if Assumption~\ref{asm-car} holds conditional
on \(A_G\) only, then \(\beta_j\) can only be consistently estimated
using \(\gamma^A_0\) and \(\alpha^A_{0j}\). The estimate from the data,
\(\hat\beta_j\), will converge to some \(\beta^*_j\).
\citet{chernozhukov2022sens} (Theorem 2 and Corollary 2) then establish
the following result.

\begin{theorem}[]\protect\hypertarget{thm-sens}{}\label{thm-sens}

When Assumption~\ref{asm-car} holds conditional on \(A_G\), then
\(|\beta^*_j - \beta_j| \le \rho S C_\gamma C_\alpha\), where \[
\begin{aligned}
\rho &:= |\mathrm{Cor}(\gamma^A_0 - \gamma_0, \alpha^A_{0j} - \alpha_{0j})|, \qquad
S^2 := \E[(\bar Y - \gamma_0(\bar X, Z))^2]\E[\alpha_{0j}(\bar X, Z)^2], \\
C_\gamma^2 &:= \frac{\E[(\gamma^A_0 - \gamma_0)^2]}{\E[(\alpha^A_{0j} - \alpha_{0j})^2]}
= R^2_{\bar Y \sim A \mid \bar X, Z}, \qand
C_\alpha^2 := \frac{\E[{\alpha^A_{0j}}^2] - \E[\alpha_{0j}^2]}{\E[\alpha_{0j}^2]}
= \frac{1 - R^2_{\alpha^A_{0j}\sim \alpha_{0j}}}{R^2_{\alpha^A_{0j}\sim \alpha_{0j}}}.
\end{aligned}
\]

\end{theorem}

In other words, the bias due to violations of Assumption~\ref{asm-car}
is bounded by the product of four terms. The first, \(\rho\), can be
upper bounded by 1, which represents adversarial confounding. It can
also be benchmarked to observed covariates, as discussed below. The
second, \(S\), is a scaling factor which can be estimated from the data.
The third, \(C_\gamma\), measures the proportion of the residual
variation in \(\bar Y_g\) explained by the unobserved confounder
\(A_G\). The fourth, \(C_\alpha\), decreases with the proportion of the
residual variation in the Riesz representer \(\alpha_{0j}\) explained by
the unobserved confounder \(A_G\).

Theorem~\ref{thm-sens} can also be directly applied to differences of
the form \(\beta_j-\beta_k\) (or, more generally, any linear contrast),
since the Riesz representer for such differences is simply
\(\alpha_{0j}-\alpha_{0k}\). When researchers are primarily interested
in differences between groups, this approach can yield tighter bounds
than applying Theorem~\ref{thm-sens} to each group separately and then
using the triangle inequality.

Researchers can vary the \emph{sensitivity parameters} \(C_\gamma\) and
\(C_\alpha\) to understand how sensitive their estimates are to
violations of Assumption~\ref{asm-car}. In fact, the entire sensitivity
analysis can be visualized on a single plot, by plotting contours of the
bound against \(C_\gamma\) and \(C_\alpha\) as contour lines. This type
of plot is familiar to causal inference researchers, who use it to
visualize sensitivity to confounding in observational studies.

As a minimal alternative to a sensitivity plot, researchers can
calculate the \emph{robustness value}, which measures the minimum
assumption violation (in terms of \(C_\gamma\) and \(C_\alpha\)) needed
to cause a bias of a specified amount. Formally, \(RV(\delta)\) is the
maximum value \(RV\) such that
\(R^2_{\bar Y \sim A \mid \bar X, Z}\le RV\) and
\(1-R^2_{\alpha^A_{0j}\sim \alpha_{0j}}\le RV\) imply
\(|\beta^*_j - \beta_j|< \delta\). In other words, if either
\(R^2_{\bar Y \sim A \mid \bar X, Z}\) and
\(1-R^2_{\alpha^A_{0j}\sim \alpha_{0j}}\) are both smaller than
\(RV(\delta)\), then the bias is less than \(\delta\). Possible values
for \(\delta\) include a certain multiple of the standard error of
\(\hat\beta_j\), or a substantively meaningful threshold. For example,
in comparing groups \(X=1\) and \(X=2\), \(RV(\hat\beta_2-\hat\beta_1)\)
would measure the minimum confounding needed to explain away the entire
estimated difference between the two groups.

For inference, \citet{chernozhukov2022sens} propose a DML estimateof the
bounds, \(\hat\beta_j \pm \hat\sigma\hat\nu|\rho|C_\gamma C_\alpha\),
where \[
\hat\sigma^2 := \E_m[(\bar Y_g - \hat\gamma(\bar X_g, Z_g))^2] \qand
\hat\nu^2 := \E_m[2\hat\alpha_j(e_j, Z_g) - \hat\alpha_j(\bar X_g, Z_g)^2],
\] Because these use the same nuisance functions \(\hat\gamma\) and
\(\hat\alpha_j\), and both estimators are based on Neyman-orthogonal
representations, these estimates will be semiparametrically efficient by
the same argument as for Theorem~\ref{thm-dml} under slightly modified
regularity conditions. The full conditions are stated in
\citet{chernozhukov2022sens}, who also propose DML confidence bounds for
these bounds which involve further computation.

\subsubsection{Interpretation}\label{interpretation}

Interpreting \(C_\gamma\) is relatively straightforward as a
(nonparametric) partial \(R^2\) of \(\bar Y_G\) on \(A_G\), conditional
on \(\bar X_G\) and \(Z_G\). Interpreting \(C_\alpha\) is more
difficult, since \(\alpha_{0j}\) is defined implicitly by
Corollary~\ref{cor-rr}. Appendix~\ref{sec-app-riesz} derives an explicit
representation of \(\alpha_{0j}\) as a weighted log derivative of the
conditional density of \(\bar X_G\) given \(Z_G\), \[
\alpha_{0j} = -u(N_g, \bar X_{gj})\partial_{\bar x_j}\log f_{\bar x\mid z}(\bar X_G, Z_G),
\] with \(\alpha^A_{0j}\) defined analogously but conditional on \(A_G\)
as well. When \(\bar X_{Gj}\mid Z_G\) is homoskedastic Gaussian, then we
have \[
\alpha_{0j}\propto N_g\bar X_{gj}(\bar X_{gj} - \E[\bar X_{gj}\mid Z_g]),
\] and if \(\E[\bar X_{gj}\mid Z_g]\) is not particularly variable
(i.e., \(R^2_{\bar X_{j}\sim Z}\) is small), then \(C_\alpha^2\) is
approximately upper bounded by
\(R^2_{\bar X_j\sim A\mid Z}/(1 - R^2_{\bar X_j\sim A\mid Z})\), which
is increasing in \(R^2_{\bar X_j\sim A\mid Z}\) So, in \emph{very} rough
terms, \(C_\alpha\) measures how much of the variation in
\(\bar X_{Gj}\) is explained by \(A_G\), conditional on \(Z_G\). In
practice, we recommend that researchers benchmark \(C_\alpha\) to
observed covariates to help in judging the plausibility of different
values of \(C_\alpha\). This benchmarking, described in
Appendix~\ref{sec-bench} and demonstrated in the application, is used in
causal inference as well.

\section{Validation}\label{sec-valid}

This section validates the proposed method in a simulation study and on
real-world data where the ground truth is known. In simulations, the
estimator outperforms alternatives, and both the global and local
confidence intervals achieve nominal coverage. In real-world data where
standard linear regression badly misses the ground truth, our method
that models a basis expansion of dozens of covariates reduces the
estimation error to within a percentage point for most groups.

\subsection{Simulation studies}\label{sec-sim}

We examine the performance of our method on data simulated from the data
generating process assumed by the now-standard method of
\citet{king1997solution}. This data-generating process draws \(\bar X\)
and \(Z\) in a correlated manner, with \(\bar X\in\Delta^d\), and then
draws \(B\) conditional on \(Z\) from a Normal distribution truncated to
the unit hypercube, so that each \(B_j\in[0, 1]\). The aggregate outcome
\(\bar Y\) is then directly calculated as \(B^\top \bar X\). For
simplicity, the size of each geography is assumed uniform, i.e.,
\(N=1\). We simulate different levels of confounding by changing both
the correlation between \(\bar X\) and \(Z\), and with the correlation
between \(B\) and \(Z\) fixed at 0.2. The entries in \(B\) are also
correlated, with a pairwise \(R^2=0.25\). Full details of the data
generating process are in Appendix~\ref{sec-valid-detail}.

In the first simulation study, we generate 1,000 datasets with \(m=500\)
geographies, \(d=2\) predictors, \(p=3\) covariates, and moderate
confounding: \(R^2_{B\sim Z}=R^2_{\bar X\sim Z}=0.5\). This 2-by-2
caseallows us to compare our method to existing methods which only
support \(d=2\). On each of the 1,000 data replicates, we applied (1)
our proposed method, including covariates entered linearly, (2) linear
regression without covariates \citep{goodman1953ecological}, (3) the
truncated-normal model of \citet{king1997solution}, from the R package
\texttt{ei}, both with and without covariates, and (4) the
Multinomial-Dirichlet count model of \citet{rosen2001bayesian},
implemented as \texttt{ei.MD.Bayes} in the R package \texttt{eiPack},
both with and without covariates.

\begin{table}

\centering{

\centering
\begin{tabular}[t]{lccccc}
\toprule
 & Covariates? & RMSE & Coverage ($50\%$) & Coverage ($95\%$) & Time (s)\\
\midrule
Proposed method & Yes & 0.013 & 0.62 & 0.99 & 0.05\\
Goodman (linear regression) &  & 0.054 & 0.11 & 0.31 & 0.01\\
Goodman (linear regression) & Yes & 0.014 & 0.41 & 0.86 & 0.00\\
King (1997; ei) &  & 0.051 & 0.08 & 0.22 & 10.00\\
King (1997; ei) & Yes & 0.016 & 0.30 & 0.74 & 83.00\\
RJKT (2002; eiPack) &  & 0.064 & 0.48 & 0.86 & 2.40\\
RJKT (2002; eiPack) & Yes & 0.083 & 0.39 & 0.75 & 9.80\\
\bottomrule
\end{tabular}

}

\caption{\label{tbl-sims}\textbf{Comparison of existing methods on
simulated data, 2×2 case}. Root mean squared error (RMSE), coverage of
nominal 50\% and 95\% confidence intervals, and average computation time
(in seconds) for different methods on simulated data. Each data
replicate contained \(m = 500\) precincts. A `X' in the covariates
column indicates that the method controlled for confounding covariates.
RJKT refers to \citet{rosen2001bayesian}.}

\end{table}%

The proposed method achieved the lowest root mean square error (RMSE) in
estimating the global parameters \(\beta\), with King's
\citeyearpar{king1997solution} model with covariates a close second.
Table~\ref{tbl-sims} presents the results. The three methods that did
not control for confounding all had similar error, around 3--4 times
higher than the proposed method. The confidence intervals for the
proposed achieved nominal coverage and in fact moderately over-covered.
None of the other methods achieved close to nominal coverage, despite
the data being drawn from a model that is exactly consistent with the
model fit by King's method. Even more concerningly, the model of
\citet{rosen2001bayesian}, which is the only method implemented in
public software that can handle \(d>2\), suffers \emph{higher} error and
\emph{lower} coverage rates when covariates are included. Finally,
estimation in competing methods is two orders of magnitude slower than
our method when covariates are not used, and even more when covariates
are included.

\begin{figure}[t]

\begin{minipage}{0.50\linewidth}

\centering{

\pandocbounded{\includegraphics[keepaspectratio]{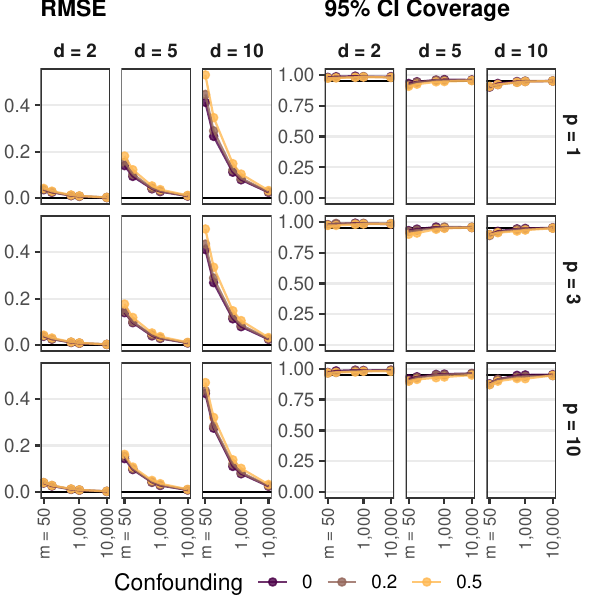}}

}

\subcaption{\label{fig-sim-seine}\textbf{Global estimates}}

\end{minipage}%
\begin{minipage}{0.50\linewidth}

\centering{

\pandocbounded{\includegraphics[keepaspectratio]{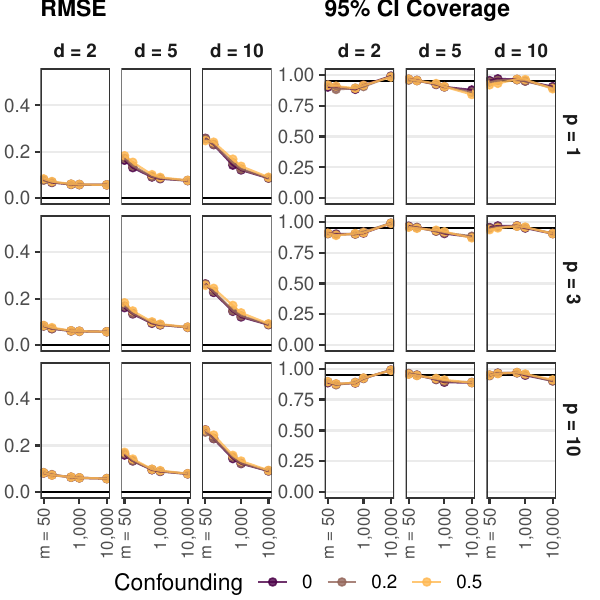}}

}

\subcaption{\label{fig-sim-local}\textbf{Local estimates}}

\end{minipage}%

\caption{\label{fig-sims}\textbf{Error and coverage on simulated data}.
RMSE and coverage of 95\% nominal confidence intervals for (a)
\textbf{global parameters} \(B\) and (b) \textbf{local parameters}
\(B_G\). Proposed method is run on 1000 replicates of simulations with
different sample sizes, numbers of predictors (columns; \(d\)), number
of covariates (rows; \(p\)), and strength of confounding (colors;
measured as the \(R^2_{\bar X\sim Z}\) with \(R^2_{B\sim Z}\) is fixed
at 0.2).}

\end{figure}%

In the second simulation study, we vary
\(m\in\{50, 100, 500, 1\,000, 10\,000\}\) (with \(m=50\) mimicking a
50-state regression), \(d\in\{2,5,10\}\), \(p\in\{1,3,10\}\), and
\(R^2_{\bar X\sim Z}\in\{0, 0.2, 0.5\}\), while fixing
\(R^2_{B\sim Z}=0.2\), with 1,000 simulated datasets for each
combination. We applied the proposed method, with covariates entering
linearly, to each simulated dataset, and also calculated local
confidence intervals using the method in Section~\ref{sec-local}.
Figure~\ref{fig-sims} shows the RMSE and coverage results for both
global and local estimates.

As our theoretical results predict, error in the global estimates
converges to \(0\) as the number of geographies increased; error in the
local estimates decreases but is lower-bounded by the intrinsic variance
of the local parameters. Error was little affected by the number of
covariates \(p\) or the strength of confounding (correlation with
\(\bar X\)), but did increase substantially with the number of
predictors \(d\). This indicates that in many ecological inference
applications, the main statistical challenge is that of many predictors,
not many covariates. This further supports the routine use of many
covariates. Across combinations of \(m\) and \(p\), coverage rates were
close to their nominal levels for \(d>2\), but above nominal levels for
\(d=2\). This is somewhat surprising given that the error grows with
\(d\). Coverage of the global and local confidence intervals is close to
the nominal level, though the coverage of the local intervals falls
somewhat below 95\% coverage for large \(m\) and \(d>2\).\footnote{ We
  observe undercoverage as well for regression-based estimates of the
  global parameter without the Riesz representer adjustment, suggesting
  finite-sample estimation error in the regression may be to blame.}

\subsection{Voter file validation}\label{sec-valid-real}

The simulation study results, while encouraging, have the virtue of a
data-generating process that exactly satisfies the required assumptions
here. We therefore turn next to a much more challenging real-world
setting, where we cannot verify that the assumptions hold exactly. This
also provides an opportunity to test the sieve estimation methods for
\(\gamma\) and \(\alpha\); in the simulation studies, the true models
were linear in the covariates.

Our data consist of 1,759 precincts in the Miami metropolitan area. The
quantity of interest is the proportion of a racial group's party
registrants who register for the Republican party. The Miami area has a
mix of different racial groups, including Cuban Americans, who are
well-known to political observers as having systematically more
Republican political preferences than other Hispanic groups, so it
serves as a good test case for our method.\footnote{ For example, the
  registration file reveals that 40 percent of Hispanic registrants
  living in Census tracts where the majority of Hispanic voters are of
  Cuban origin are Republicans, but only 24 percent of Hispanic
  registrants living in other Census tracts are Republican. This
  correlation between a covariate and the outcome of interest would lead
  to bias unless one can properly adjust for confounding covariates.}
Data on party registration come from Florida voter registration records,
and we augment this data with Census data on the racial composition of
each precinct, along with other covariates such as Hispanic origin,
population density, income, age, and past election results.
Section~\ref{sec-valid-detail} describes the voter file data and
covariates in more detail. Crucially, in Florida, voter registration
records record both a voter's party registration and their racial
affiliation, so we observe the true value of the estimand.

\begin{figure}[t]

\centering{

\pandocbounded{\includegraphics[keepaspectratio]{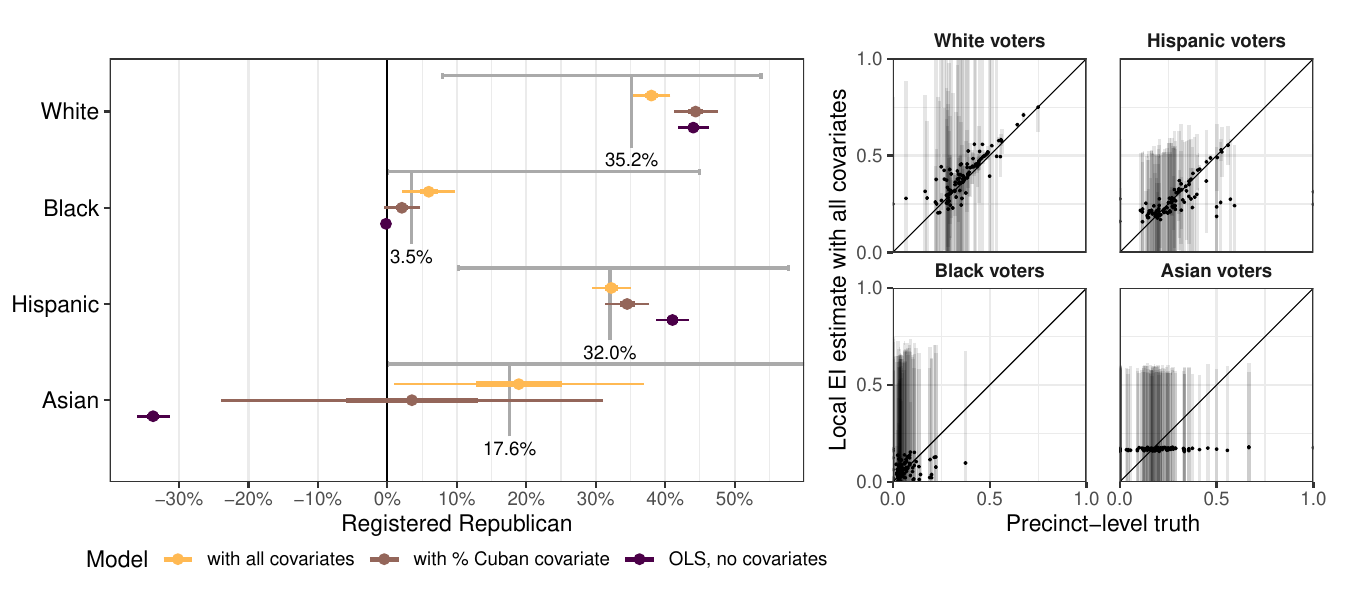}}

}

\caption{\label{fig-miami-topline}\textbf{Accuracy of predicting
Republican registration by racial group.} \textbf{(a) Global accuracy.}
The labeled vertical lines indicate the true value of Republican
registration from the voter file in the Miami metropolitan area, and the
horizontal grey lines indicate the partial identification bounds for
each group. \textbf{(b) Local accuracy.} A scatterplot of the local
estimates versus true values of Republican registration for each racial
group for 100 randomly sampled precincts. 95\% local confidence
intervals are shown for each point.}

\end{figure}%

We apply our proposed method using three different sets of covariates
for fitting \(\gamma\) and \(\alpha\). The main specification controls
for all 16 continuous covariates and dummy variables for the county and
subdivision (around 30 levels), and uses a \(J_m=1000\) tensor-product
cosine basis as described in Section~\ref{sec-bases}. A second
specification uses only one covariate, the percentage of Hispanic adults
in the Census tract that are of Cuban origin, modeled using the same
basis expansion with \(J_m=100\). Finally, we also fit our method with
no covariates and thus without penalization, which is equivalent to a
simple linear regression. The estimates from all three specifications
are displayed in the left panel of Figure~\ref{fig-miami-topline} for
the four major racial groups, along with the true values from the voter
file.

A linear regression with no covariates overestimated White GOP
registration by \(9\) percentage points (pp), overestimated Hispanic GOP
registration by \(9\)pp, and produced impossible, negative estimates for
Black and Asian voters. Controlling for covariates with the proposed
method moves all of these estimates in the correct direction. In the
more complex model with all covariates, the estimate for White voters is
only \(2\)pp off, and the estimate of Hispanic voters is only \(0.4\)pp
off. Estimates for Black voters are also no longer negative and only
\(2.5\)pp off. The estimate for Asian voters is quite variable, given
the small fraction of Asian voters in Miami, but the error is still a
double-digit improvement over the simple regression. Importantly, all
four confidence intervals for the full specification cover the true
value (just barely, for White voters).

Finally, we also evaluate the accuracy of the precinct-level local
estimates obtained using the methods in Section~\ref{sec-local}. The
right panel of Figure~\ref{fig-miami-topline} shows a scatterplot of the
local estimates versus the true values for 100 randomly sampled
precincts. For larger racial groups like White and Hispanic voters, the
local estimates are quite accurate, with an overall RMSE of \(7.3\)pp
and \(9.7\)pp, respectively. For smaller racial groups, the estimates
are shrunk towards a global mean, and the RMSE is higher: \(14\)pp for
Asian voters, for instance. Critically, however, the local confidence
intervals for all groups cover at least the nominal rate. Even using the
narrower intervals implied by a unimodality assumption, coverage of
\(95\%\) intervals (averaged across precincts) is \(96\%\), \(97.6\%\),
\(99.8\%\), and \(97.9\%\) for White, Hispanic, Black, and Asian voters,
respectively.

\section{Application}\label{sec-appl}

We now apply our method to the problem studied by \citet{jbaily2022air}:
estimating exposure to fine particulate matter (\(\text{PM}_{2.5}\)) by
racial and income groups. This application also illustrates our
sensitivity analysis.

Our data consist of 31,853 ZIP Code Tabulation Areas (ZCTA) in the U.S.
The outcome is the average \(\text{PM}_{2.5}\) exposure in 2016 for each
ZCTA, and our main predictor variable is race by income combination. The
predictors are coded as seven household income bins and two racial
groups, White and Other. Our covariates consists of ZCTA-level
population density, fraction of the over-65 population in poverty,
fraction of the over-65 population without a high-school degree, an
indicator for urbanity, as well as the latitude and longitude of the
centroid of each ZCTA. We apply the proposed estimator using a
tensor-product cosine basis on the non-geographic covariates (18 terms)
and the geographic coordinates (400 terms). All in all, the ridge
regression procedure fits 5,852 coefficients; this takes around 30
seconds on a modern laptop.

As discussed in Section~\ref{sec-est-semi}, the second moments of the
fitted Riesz representer serve as a way to assess the positivity
assumption which underlies estimation. Here, they range from \(26.1\) to
\(74.9\), which is larger than the minimum possible value of 1. This
reflects the fact that few ZCTAs comprise entirely one race-income
group: most of the \(\bar X_j\) are closer to 0 than to 1. As a result,
more extrapolation is needed to estimate the conditional mean for each
group. Exploratory analysis confirms reasonable variation in each
\(\bar X_j\), however, and so we believe that Assumption~\ref{asm-pos}
is plausible. The larger second moments will lead to more variable
estimates, however.

\begin{figure}[t]

\centering{

\pandocbounded{\includegraphics[keepaspectratio]{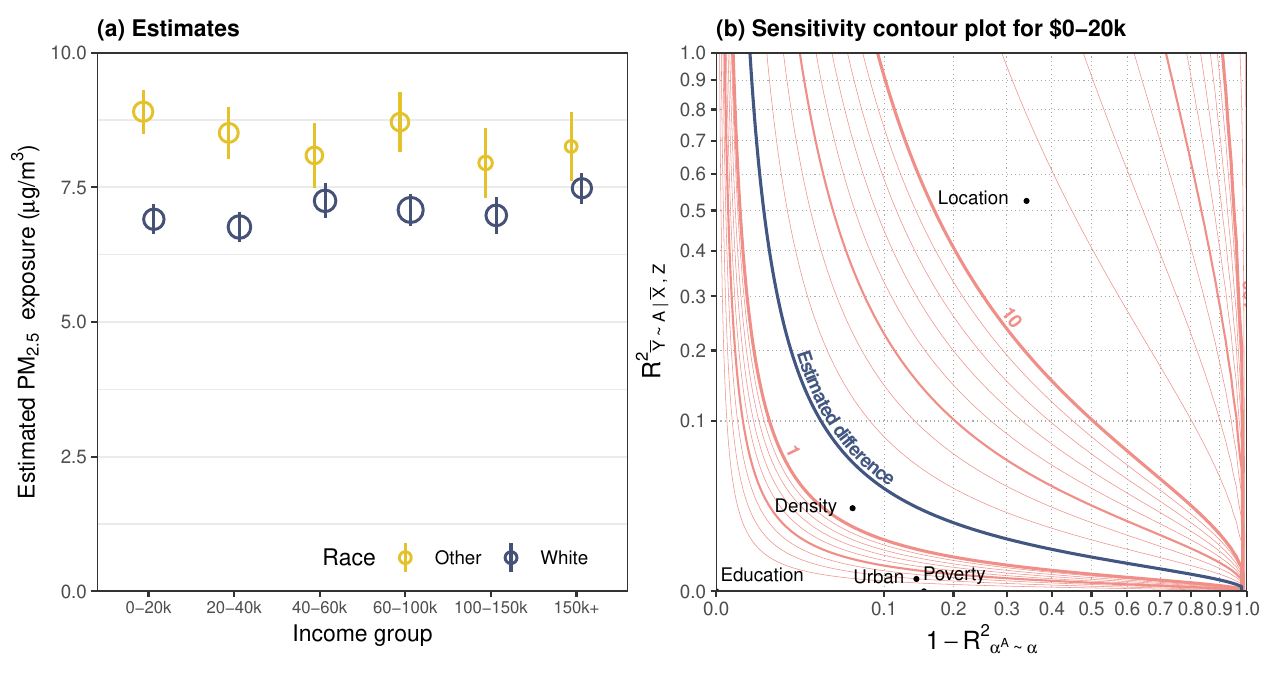}}

}

\caption{\label{fig-air}\textbf{(a) Estimates of pollution exposure by
race and income group.} Circles are centered at the estimate and have
area proportional to the size of each group in the U.S. population.
Vertical lines display 95\% confidence intervals. \textbf{(b)
Sensitivity analysis for the racial disparity in exposure among people
earning less than \$20,000.} The two sensitivity parameters plotted
along each axis, and the contours indicate the bias in the estimated
difference (Other - White) that would arise from the specified degree of
confounding. The blue contour corresponds to bias that would be
sufficient for the estimated disparity to be zero. Benchmarked
sensitivity parameters for observed covariates are also plotted.}

\end{figure}%

Figure~\ref{fig-air} (a) displays the estimates for each race and income
group along with 95\% confidence intervals. There is no clear income
disparity within either racial group, but there are clear disparities
across racial groups within the lower income categories. These
disparities are statistically significant but not particularly large:
monthly variation in \(\text{PM}_{2.5}\) exposure can be on the order of
\(10\ \mu g/m^3\) \citep{rao2011understanding}. The direction of this
disparity is consistent with the findings of \citet{jbaily2022air}.

The estimates in Figure~\ref{fig-air} (a) rely on
Assumption~\ref{asm-car} holding: that conditional on the population
density, education, poverty, urbanity, and approximate geographic
location of a ZCTA, air pollution exposure is unrelated to the racial
composition of the ZCTA. While this assumption seems plausible,
especially due to the control for geographic location, it is important
nonetheless to assess the sensitivity of the estimates to violations of
this assumption.

For simplicity, we show only the sensitivity analysis for the difference
in exposure between Other and White residents earning less than \$20,000
per year (leftmost points in Figure~\ref{fig-air} (a)). The point
estimate is that the non-White population is exposed to \(2\ \mu g/m^3\)
more pollution than the White population. We first calculate the
robustness value for bias equal to this point estimate, which is
\(0.0314\). This means that if either
\(R^2_{\bar Y\sim A\mid \bar X, Z}\) or \(1-R^2_{\alpha^A\sim \alpha}\)
is larger than this value, then the bias could be large enough to
explain away the entire estimated disparity.

To better interpret these robustness values, Figure~\ref{fig-air} (b)
shows a sensitivity contour plot. For each combination of sensitivity
parameters, the contour lines indicate the size of the bias that would
arise from confounding of that magnitude. The contour labeled
``Estimated difference'' marks the dividing line at which the disparity
estimate would change sign. This contour is rather close to the origin,
indicating substantial sensitivity, which agrees with the high
sensitivity implied by the small robustness value.

Figure~\ref{fig-air} (b) also displays benchmarked values of the
sensitivity parameters for observed covariates (detailed in
Appendices~\ref{sec-bench} and \ref{sec-pollution}). These benchmarks
show that if an omitted confounder is of similar strength to ZCTA
education, population density, urbanity, or poverty, it would likely not
change the sign of the disparity estimate.

In contrast, the location variable has a much larger benchmarking value.
The value is closer to 40, which is far larger than the estimated
difference. In other words, if the omitted confounder is of similar
strength to geographic location, then it would easily change the sign of
the estimate and create substantial bias. In a more in-depth analysis,
these findings would prompt us to consider collecting other covariates,
and more carefully evaluate the model specification as regards the
critical geography covariate.

\section{Conclusion}\label{sec-concl}

We have formalized the identification assumptions for ecological
inference and proposed a new set of tools for estimation and sensitivity
analysis. We stress for practitioners the importance of thinking
carefully about the identifying assumptions, rather than blindly
applying existing ecological inference methods without any covariates.
With the tools presented here, the plausibility of and sensitivity to
these identifying assumptions can be directly assessed, and when they
are judged reasonable, estimation may be carried out efficiently. We
strongly recommend the routine use of sensitivity analysis in performing
ecological inferences.

One drawback of the proposed estimator is that when \(Y\) is bounded,
the regression \(\hat\gamma\) can be fit to respect these bounds, but
the overall estimate \(\hat\beta\) may not, due to the form of the
efficient influence function (Eq.~\ref{eq-neyman}). Future work could
explore ways to modify the estimator to respect these bounds, which may
further reduce error in finite samples.

There are other possible extensions of the methods proposed here. One
interesting case is when only \(Y\) but not \(X\) is aggregated, such as
when a voter's ballot \(Y\) is secret and can only be observed in
aggregate at the precinct level, but many individual-level covariates
\(X\) are available from voter files or surveys.
\citet{flaxman2015supported} and \citet{fishman2024estimating} have
proposed methods for this setting, but results on identification, and
estimation guarantees, remain limited. Related to this case is the
challenge of estimating conditional means for a variety of \(X\) at once
(e.g., race and income and education), or for a high-dimensional \(X\).
Another possible extension is to leverage spatial correlation in the
data, which is likely to be present in many applications (such as in
Section~\ref{sec-appl}), and may allow for estimating a latent spatial
confounder. Finally, we have focused on estimating conditional means
here (and conditional variances in Section~\ref{sec-local}), but other
estimands may be of interest, such as quantiles.

\section*{References}\label{references}
\addcontentsline{toc}{section}{References}

\renewcommand{\bibsection}{}
\bibliography{references.bib}

\newpage{}

\appendix

\section{Testing the Coarsening at Random Assumption}\label{sec-id-test}

In causal inference, sensitivity analyses are often motivated by the
fact that causal identifying assumptions are not testable in the data.
The same might appear to be true in ecological inference, but as we have
seen, Assumption~\ref{asm-car} actually implies that the CEF
\(\E[\bar Y_G\mid \bar X_G, Z_G]\) is partially linear in \(\bar X_G\).
This implication is in principle testable, opening a route to testing
Assumption~\ref{asm-car} in real-world data.

\subsection{Method}\label{method}

Specifically, by testing a fully nonparametric regression of
\(\bar Y_G\) on \(\bar X_G\) and \(Z_G\) against the partially linear
regression of \(\bar Y_G\) on \(\bar X_G\) and \(Z_G\) used in the main
estimation routine, we can test the null hypothesis that
Assumption~\ref{asm-car} holds. We propose augmenting the partially
regression model with additional sieve basis functions applied to
\((\bar X_G, Z_G)\) rather than just \(Z_G\). In this way, the
functional form test becomes an omnibus test of the coefficients on
these additional terms.

Carrying out the test is complicated by the ridge penalization.
Following \citet{helwig2022robust}, we use a Kennedy--Cade
\citeyearpar{kennedy1996randomization} permutation test using the Wald
test statistic, which Helwig showed is superior to an F statistic. The
Wald statistic is the squared norm of the tested coefficients under the
metric given by the covariance matrix of those coefficients, \[
W := \hat\beta^\top \hat\Sigma_\beta^{-1} \hat\beta,
\] where \(\hat\Sigma_\beta\) is the estimated covariance matrix. Helwig
recommends using a heteroskedasticity-consistent covariance matrix
estimator, and we have found this to be important in simulations, though
it does increase the computational cost of the procedure.

The test is carried out by first residualizing both \(\bar Y_G\) and the
additional sieve basis functions on the restricted regression terms,
\(\bar X_G\otimes \Phi_m(Z_G)\). We have found in simulation studies
that slightly undersmoothing during this residualization step (i.e.,
using a smaller penalty than the one selected by LOOCV by a factor of
\(1.5\)--\(2\) or so) can improve Type I error control. After
residualizing, the permutation test is carried out by permuting the
residualized \(\bar Y_G\) and calculating the Wald test statistic for
the coefficients on the residualized additional sieve basis functions
for each permutation.

In large samples, the asymptotic distribution of the Wald statistic
under the null hypothesis is \(\chi^2_q\), where \(q\) is the number of
additional sieve basis functions. This approximation may be used to
reduce the computational cost of the test in large samples. The
simulation study below suggests the approximation is accurate enough
once \(n > 1\,000\) or so.

The proposed test has two main drawbacks. First, both Type I and Type II
error rates can be sensitive to the choice of penalty in the first
residualization step, and so in finite samples the test may not be valid
or may have very low power, without the researcher knowing. Second,
rejection of the test does not provide a measure of the degree of
violation of Assumption~\ref{asm-car}, and so does not directly inform
the likely amount of bias present. Thus we still recommend that
researchers always conduct a sensitivity analysis, which directly
connects the size of the assumption violation to the amount of bias,
regardless of the power of the test proposed here.

Both of these drawbacks can be ameliorated by generating semi-synthetic
data with a known confounder and a range of degrees of confounding,
informed by the sensitivity analysis. Applying the proposed test to
these various semi-synthetic data sets can identify inflated Type I
error and give researchers a sense of the power of the test. In a
best-case scenario, researchers might find that a confounder strong
enough to lead to a problematic amount of bias would also lead to
rejection of the test with high probability. If in fact the test does
not reject, this would provide a strong indication that the bias is in
fact small.

\subsection{Validation}\label{validation}

We evaluate our proposed test of Assumption~\ref{asm-car}. We consider a
subset of the simulations in Study 2 where \(d=2\), \(p=3\),
\(R^2_{\bar X\sim Z}=0.5\), and \(R^2_{B\sim Z}=0.2.\) For computational
reasons, we only use \(200\) of the \(1\,000\) replicates for
\(m=10\,000\). These simulations are generated under CAR. We consider
estimating the \(p\)-value associated with the null hypothesis with the
permutation test as well as using the \(\chi^2\) approximation, and we
consider specifications using the LOO-selected ridge regression penalty
as well as an undersmoothed penalty (half the LOO-selected penalty).

\begin{figure}[t]

\centering{

\pandocbounded{\includegraphics[keepaspectratio]{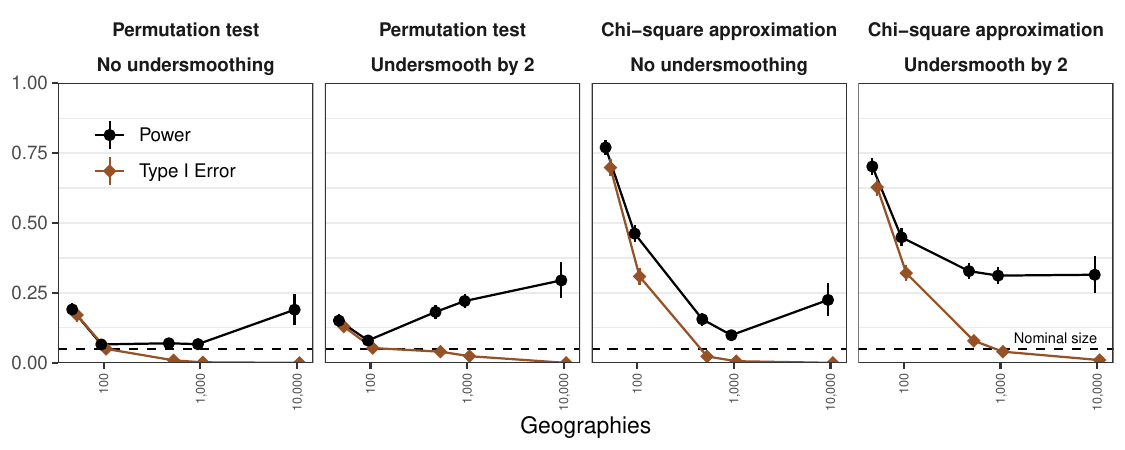}}

}

\caption{\label{fig-sim-car-test}\textbf{Power and Type I error of the
proposed test of Assumption~\ref{asm-car}}. Power is calculated under
the alternative hypothesis where one of the covariates used to generate
the data is omitted from model fitting. 95\% confidence intervals are
shown for each estimate of power and Type I error. The four panels vary
the method of estimating the \(p\)-value (permutation test
vs.~\(\chi^2\) approximation) and the choice of ridge penalty
(LOO-selected vs.~undersmoothed).}

\end{figure}%

Figure~\ref{fig-sim-car-test} shows the power and Type I error of the
test across the 5 different sample sizes and the four test
configurations. Power is calculated under the alternative hypothesis
where one of the covariates used to generate the data is omitted from
model fitting. We consider this to be a moderately low-signal
alternative, since the overall amount of confounding is
small-to-moderate, and only one of three covariates is omitted.

Except at the smallest sample size \(m=50\), the test's error rate is
below the nominal size, with power increasing with sample size.
Undersmoothing improves power and Type I error control across sample
sizes. We find that the \(\chi^2\) approximation (the third and fourth
panels) is accurate enough for \(m\ge\) 1,000 or so, but is highly
anti-conservative at smaller sample sizes, as expected. Overall, we
recommend using the more computationally intensive permutation test
where possible, and we caution that the test may have lower power in
certain settings.

\section{Practical Implementation Details}\label{sec-app-impl}

\subsection{Bounded outcomes}\label{bounded-outcomes}

When \(Y\) is an indicator variable (or generally when it is bounded),
then \(\bar Y\) is bounded, and so the predictions for each group, i.e.,
\((e_j\otimes \phi_m(Z_g))^\top \theta\), should be bounded as well.
These bounds can be expressed as linear constraints in the optimization
problem in Eq.~\ref{eq-sieve-est}. Since the criterion is quadratic, the
resulting optimization problem is a quadratic program, which can be
solved almost as efficiently as the unconstrained problem. Of course,
the same idea can be applied when \(Y\) is bounded on one side only. We
implement this approach in our software. Note, however, that while
bounding \(\hat\gamma_m\) can reduce variance, it does not guarantee
that the final estimate \(\hat\beta_{mj}\) will respect the same bounds,
due to the second (correction) term in Eq.~\ref{eq-neyman}, which may be
negative.

\subsection{Weighted estimation}\label{weighted-estimation}

As Theorem~\ref{thm-id} shows, Assumption~\ref{asm-car} is sufficient
for identification of \(\beta_j\) and estimation without weighting by
\(N_g\), as has been suggested by previous work. However, weighting can
still be useful for finite-sample efficiency when there is
heteroskedasticity in \(\bar Y_g\). Sometimes, the variance of
\(\bar Y_g\) may scale inversely with \(N_g\), which would justify
weighting by \(N_g\) from an efficiency perspective rather than an
identification perspective. But in many cases, including in analyses of
voting data, there is no particular reason to assume this form for the
variance. Practitioners' substantive knowledge may be used to construct
approximate inverse-variance weights that can reduce estimation variance
in finite samples.

\subsection{Possible sieve bases}\label{sec-bases}

We provide some examples of sieve bases and function classes which
satisfy Assumption~\ref{asm-sr} (4)--(6), and the resulting rates that
are achieved in Theorem~\ref{thm-sieve}. That is, estimates of
\(\gamma_0\) and \(\alpha_{0j}\) converging at a rate faster than
\(m^{-1/4}\) (where \(m\) is the number of geographies) will be
sufficient for \(\hat\beta_{mj}\) to be asymptotically normal and
semiparametrically efficient.

For simplicity we take \(\supp(Z_g)=[0,1]^p\). Rather than formally
defining each function class and basis, we provide references to the
relevant literature for full details.

\begin{itemize}
\item
  \emph{Polynomials}. If \(\F=\H^q\), the Hölder space of \(q\)-smooth
  functions, consider \(\Phi_m\) being the set of polynomials where each
  variable (not term) has degree at most \(J_m\), i.e., all \(p\)-way
  interactions of polynomials in each variable up to degree \(J_m\). In
  this setting, \(A_m\asymp J_m^p\) (meaning \(A_m=O(J_m^p)\) and
  \(J_m^p=O(A_m)\)) and \(\rho_m=O(J_m^{-q})\). If \(q>p\) then
  Assumption~\ref{asm-sr} (4)--(6) hold for the optimal sieve size of
  \(J_m\asymp m^{1/(2q+p)}\) \citep[p.~5570 \emph{et
  seq}]{chen2007large}. The resulting rate from Theorem~\ref{thm-sieve}
  is \(O_{\Pr}(m^{-q/(2q+p)})\), which is minimax optimal for estimation
  in \(\H^q\), and since \(q>p\), is in particular
  \(o_{\Pr}(m^{-1/3})\).
\item
  \emph{Tensor-product splines}. If \(\F=\H^q\) and \(\Phi_m\) is the
  set of tensor-product splines of order \(r\) (degree \(r-1\)) on
  \(J_m\) equally-spaced knots in each dimension, then
  \(A_m\asymp J_m^{p/2}\) and \(\rho_m=O(J_m^{-q})\). If \(q>p/2\) and
  \(r\ge\lfloor q\rfloor+1\), then Assumption~\ref{asm-sr} (4)--(6) hold
  for the optimal sieve size of \(J_m\asymp m^{1/(2q+p)}\)
  \citep{chen2007large}. The resulting rate from Theorem~\ref{thm-sieve}
  is again \(O_{\Pr}(m^{-q/(2q+p)})\), and since \(q>p/2\), is in
  particular \(o_{\Pr}(m^{-1/4})\).
\item
  \emph{Tensor-product cosine basis}. Let
  \(\F=S_1([0,1]^p)\cong\bigotimes_{i=1}^p W_1([0, 1])\), the
  first-order Sobolev space with dominating mixed derivatives, which is
  identified with the tensor product space of univariate first-order
  Sobolev spaces \citep{zhang2023regression}. Those authors show that if
  \(\Phi_m\) is a set of \(J_m\) tensor products of univariate cosine
  basis functions \(\phi_j(z) = \sqrt{2}\cos(\pi(j-1)z)\), with the
  products included in the sieve in a certain order, then letting
  \(J_m\asymp m^{1/3}(\log m)^{2(p-1)/3}\) is optimal and yields a rate
  of \(O_{\Pr}(m^{-1/3}(\log m)^{(p-1)/3}\log m)\), which is a
  \(\log m\) factor away from minimax optimal for estimation in
  \(S_1([0,1]^p)\). However, the dependence on \(p\) in this rate is far
  better, and in particular the rate is \(o_{\Pr}(m^{-1/4})\) for all
  \(p\).
\end{itemize}

Thus, any of these options will yield consistent estimates of
\(\gamma_0\) and \(\alpha_{0j}\) at a rate sufficient for
\(\hat\beta_{mj}\) to be asymptotically normal and semiparametrically
efficient.

\subsection{Benchmarking sensitivity parameters}\label{sec-bench}

In real-world applications, researchers may have some idea of plausible
confounders \(A_G\) but may not be as sure of reasonable values for
\(\rho\), \(C_\gamma\), \(C_\alpha\). One aid to interpreting these
parameters is to benchmark them against observed covariates. This is
accomplished by leaving each covariate in turn \(Z_{Gk}\) out of the
estimation process, and then calculating \(\rho\), \(C_\gamma\),
\(C_\alpha\) as if \(A_G=Z_{Gk}\), and making an additional assumption
to account for the effect of leaving out covariates (Appendix D of
\citet{chernozhukov2022sens} describes this benchmarking in detail).
This will produce \(p\) benchmark values for each sensitivity parameter,
one for each covariate. The change in the estimate itself can also be
calculated for each left-out covariate.

It is important to stress that benchmarked sensitivity parameters
themselves do not imply any particular bounds on \(\rho\), \(C_\gamma\),
or \(C_\alpha\) for the actual confounder \(A_G\). Researchers would
need to make a strong exchangeability-type assumption between \(A_G\)
and the observed covariates to make such an inference. Despite these
limitations, benchmarking can still help researchers use their
substantive knowledge about well-known confounders to calibrate the
sensitivity parameters. We expect this to be particularly useful for
\(C_\alpha\), which is more difficult to directly interpret than
\(C_\gamma\).

\appendix

\renewcommand\thefigure{\thesection.\arabic{figure}}

\setcounter{figure}{0}

\section{Riesz Representer}\label{sec-app-riesz}

In this section, we drop the subscript \(G\) for clarity.

\subsection{Interpretation}\label{interpretation-1}

Recall that the estimand can be written as
\(\beta_j=\E[\gamma(e_j, Z)u(N,\bar X_j)]\). Because
\(\gamma(e_j, Z)=\eta_0(Z)^\top \bar X\) (Eq.~\ref{eq-cef}), we can
equivalently write the estimand as \[
\beta_j = \E[u(N,\bar X_j)e_j^\top\partial_x\gamma(\bar X, Z)],
\] i.e., the (weighted) partial derivative of \(\gamma\) with respect to
\(\bar X_j\). \citet{chernozhukov2022sens} show that the Riesz
representer of functionals of this type can be expressed as \[
\alpha_j(n, \bar x, z)
= -\frac{u(n,\bar x_j)\partial_{x_j} f(\bar x\mid z)}{f(\bar x\mid z)}
= -u(n,\bar x_j)\partial_{\bar x_j} \log f(\bar x\mid z).
\]

We can further simplify \(\alpha_j\) in some cases. When
\(\bar X_j\mid Z\) is Gaussian with homoskedastic variance
\(\varsigma_j^2=\Var[\bar X_j\mid Z]\) and all \(N=1\), then \[
\begin{aligned}
\partial_{x_j} \log f(\bar x\mid z)
&= \partial_{x_j} \qty(-\half \log (2\pi\varsigma_j^2)
- \half(\bar x_j - \E[\bar X_j\mid Z=z])^2 / \varsigma_j^2) \\
&= -\varsigma_j^{-2}(\bar x_j - \E[\bar X_j\mid Z=z]).
\end{aligned}
\] The Riesz representer in this case is therefore \[
\alpha_j(n,\bar x,z) = \varsigma_j^{-2}\E[\bar X_j]^{-1}\bar x_j(\bar x_j - \E[\bar X_j\mid Z=z]).
\] For the extended Riesz representer, the conditioning would
additionally be on the unobserved confounder \(A\), and \(\Omega\) would
change. The sensitivity parameter \(1-R^2_{\alpha\sim\alpha_s}\) has a
simple interpretation in this case as well. We have \[
\begin{aligned}
\E[\alpha_j^2] &= \varsigma_j^{-4}\E[\bar X_j]^{-2}\E\qty[\bar X_j^2(\bar X_j - \E[\bar X_j\mid Z])^2] \\
&= \varsigma_j^{-4}\E[\bar X_j]^{-2}\varsigma_j^{2}(\E[\E[\bar X_j\mid Z]^2]+3\varsigma_j^{2}) \\
&= \varsigma_j^{-2}\frac{\E[\E[\bar X_j\mid Z]^2]}{\E[\E[\bar X_j\mid Z]]^{2}} + 3\E[\bar X_j]^{-2}.
\end{aligned}
\] by a simple calculation based on the moments of a Gaussian. When
\(\bar X_j\mid Z\) is not particularly variable, the Jensen gap is not
large, and we will have \(\E[\alpha_j^2]\approx \varsigma_j^{-2}\). Then
\[
C_\alpha^2 = \frac{\E[\alpha_j^2]-\E[{\alpha^A_j}^2]}{\E[\alpha_j^2]}
\approx \frac{\Var[\bar X_j\mid Z] - \Var[\bar X_j\mid Z, A]}{\Var[\bar X_j\mid Z] + 3\E[\bar X_j]^{-2}}
< \frac{\Var[\bar X_j\mid Z] - \Var[\bar X_j\mid Z, A]}{\Var[\bar X_j\mid Z]}
= \frac{R^2_{\bar X_j\sim A\mid Z}}{1 - R^2_{\bar X_j\sim A\mid Z}},
\] with the bound reasonably tight as long as
\(3\E[\bar X_j]^{-2}\ll \Var[\bar X_j\mid Z]\).

\subsection{Closed-form solution}\label{closed-form-solution}

We can express the specific \(\alpha\) here as \[
\alpha = (\bar X\otimes \Phi(Z))^\top \theta
\] for some parameter \(\theta\), so \[
m_j(\alpha) = u(N, \bar X_j)(e_j\otimes\Phi(Z))^\top \theta.
\] We will minimize a penalized version of Eq.~\ref{eq-min-alpha}: \[
L(\theta) = \E[\theta^\top (\bar X\otimes \Phi(Z))(\bar X\otimes \Phi(Z))^\top \theta]
- 2\E[u(N, \bar X_j)(e_j\otimes\Phi(Z))^\top \theta] + \lambda\theta^\top\theta;
\] a value of \(\lambda=0\) corresponds to the population criterion that
recovers \(\alpha_0\). For any \(\lambda\), we have \[
\partial_\theta L(\theta) = 2\E[(\bar X\otimes \Phi(Z))(\bar X\otimes \Phi(Z))^\top]\theta
- 2u(N, \bar X_j)(e_j\otimes\Phi(Z)) + 2\lambda\theta;
\] the second derivative is
\(2\E[(\bar X\otimes \Phi(Z))(\bar X\otimes \Phi(Z))^\top]+2\lambda\),
so the problem is convex as long as the Gram matrix is well behaved.
Thus the unique solution is given by solving
\(\partial_\theta L(\theta)=0\), yielding \[
\theta^* = (\E[(\bar X\otimes \Phi(Z))(\bar X\otimes \Phi(Z))^\top]+ \lambda I)^{-1}u(N, \bar X_j)(e_j\otimes\Phi(Z))
\] and \[
\alpha^*(N, \bar X, Z) = (\bar X\otimes \Phi(Z))^\top(\E[(\bar X\otimes \Phi(Z))(\bar X\otimes \Phi(Z))^\top]+ \lambda I)^{-1}u(N, \bar X_j)(e_j\otimes\Phi(Z)).
\]

The argument goes through identically if \(\E\) is replaced with an
empirical average \(\E_n\). Let \(\XZ\) be the matrix with each row
\(\bar X_i\otimes \Phi(Z_i)\), and \(\XZj\) the matrix with each row
\(u(N_i, \bar X_{ij})(e_j\otimes\Phi(Z_i))\). Then the finite-sample
solution is \[
\theta^* = (\XZ^\top\XZ + \lambda I)^{-1} \XZj^\top \mathbf{1}
\qand
\alpha^* = \XZ (\XZ^\top\XZ + \lambda I)^{-1} \XZj^\top \mathbf{1},
\] where now \(\alpha^*\) is a vector of ``fitted values'' for each
observation. If we use the SVD of \(\XZ=UDV^\top\), then these
expressions simplify as \[
\theta^* = V(D^2+\lambda I)^{-1}V^\top \XZj^\top \mathbf{1}
\qand
\alpha^* =  UD(D^2+\lambda I)^{-1}V^\top \XZj^\top \mathbf{1}.
\] Compare these to the expressions for ridge regression on a design
matrix with SVD \(UDV^\top\): \[
\theta^* = VD(D^2+\lambda I)^{-1}U^\top\vb{y} \qand
\hat{\vb{y}} = UD^2(D^2+\lambda I)^{-1}U^\top\vb{y}.
\]

\subsection{Value of criterion at
minimizer}\label{value-of-criterion-at-minimizer}

Our sensitivity analysis requires estimating
\(\nu=\E[2m_j(\alpha) - \alpha^2 ]\). Substituting the solutions above
using the SVD of \(\XZ\), we obtain \[
\begin{aligned}
\hat\nu &= \E_n[2m_j(\alpha^*) - \alpha^{*2}] \\
&= n^{-1}\qty(
 2\cdot \mathbf{1}^\top \XZj V(D^2+\lambda I)^{-1}V^\top \XZj^\top \mathbf{1}
 -\mathbf{1}^\top \XZj V (D^2+\lambda I)^{-1}DU^\top
 UD(D^2+\lambda I)^{-1}V^\top \XZj^\top \mathbf{1} ) \\
&= n^{-1}\cdot \mathbf{1}^\top \XZj V
    \qty(2(D^2+\lambda I)^{-1} - D^2(D^2+\lambda I)^{-2})V^\top
    \XZj^\top \mathbf{1};
\end{aligned}
\] this can be computed quickly. When \(\lambda=0\) notice that this
simplifies to \[
\hat\nu =  n^{-1}\cdot \mathbf{1}^\top \XZj V D^{-2} V^\top \XZj^\top \mathbf{1}.
\]

\subsection{Leave-one-out expressions}\label{leave-one-out-expressions}

We can write the leave-one-out solution for \(\theta\) as \[
\theta^*_{(-i)} = (\XZ^\top\XZ - \xz_i\xz_i^\top + \lambda I)^{-1} (\XZj^\top \mathbf{1} - \xzt_{ji}),
\] where \(\xz_i\) is the \(i\)-th row of \(\XZ\) and \(\xzt_{ji}\) is
the \(i\)-th row of \(\XZj\). Applying the Sherman--Morrison identity,
we have \[
\theta^*_{(-i)} = \qty((\XZ^\top\XZ + \lambda I)^{-1} +
\frac{(\XZ^\top\XZ + \lambda I)^{-1}\xz_i\xz_i^\top(\XZ^\top\XZ + \lambda I)^{-1}}{
1 - \xz_i^\top(\XZ^\top\XZ + \lambda I)^{-1}\xz_i }) (\XZj^\top \mathbf{1} - \xzt_{ji}).
\]

Then, letting here \(H=\XZ (\XZ^\top\XZ + \lambda I)^{-1} \XZ^\top\)
with diagonal entries \(h_{ii}\), the leave-one-out prediction for
\(\alpha^*_i\) is \[
\begin{aligned}
\alpha^*_{(-i)} &= \xz_i^\top\qty((\XZ^\top\XZ + \lambda I)^{-1} +
    \frac{(\XZ^\top\XZ + \lambda I)^{-1}\xz_i\xz_i^\top(\XZ^\top\XZ + \lambda I)^{-1}}{
    1 - \xz_i^\top(\XZ^\top\XZ + \lambda I)^{-1}\xz_i }) (\XZj^\top \mathbf{1} - \xzt_{ji}) \\
&= \qty(\xz_i^\top(\XZ^\top\XZ + \lambda I)^{-1} +
    \frac{\xz_i^\top(\XZ^\top\XZ + \lambda I)^{-1}\xz_i\xz_i^\top(\XZ^\top\XZ + \lambda I)^{-1}}{
    1 - \xz_i^\top(\XZ^\top\XZ + \lambda I)^{-1}\xz_i }) (\XZj^\top \mathbf{1} - \xzt_{ji}) \\
&= \qty(\xz_i^\top(\XZ^\top\XZ + \lambda I)^{-1} +
    \frac{h_{ii}\xz_i^\top(\XZ^\top\XZ + \lambda I)^{-1}}{1 - h_{ii}})
    (\XZj^\top \mathbf{1} - \xzt_{ji}) \\
&= \qty(1 + \frac{h_{ii}}{1 - h_{ii}}) \xz_i^\top(\XZ^\top\XZ + \lambda I)^{-1}
    (\XZj^\top \mathbf{1} - \xzt_{ji}) \\
&= \frac{1}{1 - h_{ii}}(\xz_i^\top(\XZ^\top\XZ + \lambda I)^{-1}\XZj^\top \mathbf{1}
    -\xz_i^\top(\XZ^\top\XZ + \lambda I)^{-1}\xzt_{ji}) \\
&= \frac{1}{1 - h_{ii}}(\alpha^*_i - \xz_i^\top(\XZ^\top\XZ + \lambda I)^{-1}\xzt_{ji}).
\end{aligned}
\] This can be expressed with the SVD of \(\XZ\) as \[
\alpha^*_{(-i)} = \frac{1}{1 - h_{ii}}\qty(\alpha^*_i - \vb{u}_iD(D^2+\lambda I)V^\top \xzt_{ji}),
\] where \(\vb{u}_i\) here is the \(i\)-th row of \(U\) (and, in an
unfortunate clash of notation, unrelated to the weighting function
\(u(N, \bar X_j)\)).

We can also calculate the leave-out-one \(m(\alpha^*)\) as \[
\begin{aligned}
m_j^{(-i)}(\alpha^*) &= \xzt_{ji}^\top(\XZ^\top\XZ + \lambda I)^{-1}
  \qty(1 + \frac{\xz_i\xz_i^\top(\XZ^\top\XZ + \lambda I)^{-1}}{1 - h_{ii}})
  (\XZj^\top \mathbf{1} - \xzt_{ji}).
\end{aligned}
\]

\section{Proofs}\label{sec-app-proofs}

\subsection{\texorpdfstring{Proof of
Theorem~\ref{thm-id}}{Proof of Theorem~}}\label{proof-of-thm-id}

\begin{proof}
As noted in the main text, Assumption~\ref{asm-car} implies
\(\E[B_G\mid \bar X_G=\bar x, Z_G=z]=E[B_G\mid Z_G=z]\) for all
\(\bar x\) and \(z\) by integrating out \(N_G\). Applying this property
once to drop the conditioning on \(\bar X_{Gj}=1\) (which, by the
sum-to-1 constraint, fixes \(\bar X_G\)) and applying
Assumption~\ref{asm-car} to condition on \(\bar X_G\) and \(N_G\), \[
\begin{aligned}
\E[N_G\bar X_{Gj}\E[\bar Y_G\mid Z_G, \bar X_{Gj}=1]]
&= \E[N_G\bar X_{Gj} \E[B_G^\top \bar X_G\mid Z_G, \bar X_{Gj}=1]] \\
&= \E[N_G\bar X_{Gj} \E[B_{Gj} \mid Z_G, \bar X_{Gj}=1]] \\
&= \E[N_G\bar X_{Gj} \E[B_{Gj} \mid Z_G]] \\
&= \E[N_G\bar X_{Gj} \E[B_{Gj} \mid Z_G, \bar X_G, N_G]] \\
&= \E[\E[N_G\bar X_{Gj} B_{Gj} \mid Z_G, \bar X_G, N_G]] \\
&= \E[N_G\bar X_{Gj} B_{Gj}] = E[N_G\bar X_{Gj}]\beta_j.
\end{aligned}
\] Dividing by \(\E[N_G\bar X_{Gj}]\) yields the result for
Assumption~\ref{asm-car}. The result for Assumption~\ref{asm-car-u}
follows by an identical argument with \(G^n\) substituted for \(G\) and
\(N_G\) dropped.
\end{proof}

\subsection{\texorpdfstring{Proof of
Theorem~\ref{thm-car-ind-agg}}{Proof of Theorem~}}\label{proof-of-thm-car-ind-agg}

\begin{proof}
For each level \(x\in\X\) and geography \(g\in\G\), \[
\begin{aligned}
\E[B_{gj}\mid Z_g, \bar X_g]
&= \E\left[ \frac{N_g^{-1}\sum_i X_{xi}Y_i\ind\{G_i=g\}}{\bar X_{gj}}
    \,\middle|\, Z_g, \bar X_g\right] \\
&= \bar X_{gj}^{-1} \E\left[\sum_i N_g^{-1}\ind\{G_i=g\} X_{ij}
    \E[Y_i\mid X_i, Z_g, \bar X_g, G_i]
    \,\middle|\, Z_g, \bar X_g\right]. \\
\end{aligned}
\] We can rewrite the inner expectation as
\(\E[Y_i\mid Z_{G_i}, X_i,  G_i]\), since
\(Y_i\indep \bar X_i\mid Z_{G_i}, X_i\) by independence. We can replace
\(Z_g\) with \(Z_{G_i}\) since this expectation is multiplied by
\(\ind\{G_i=g\}\); only when \(g=G_i\) does the value of the expectation
matter. Then applying Assumption~\ref{asm-car-ind}, we can further
simplify, yielding overall \[
\E[Y_i\mid X_i, Z_g, \bar X_g,  G_i]
= \E[Y_i\mid Z_{G_i}, X_i]
\] inside the expression above. Since \(X_{ij}\) is also an indicator
variable, we can condition on \(X_{ij}=1\) rather than \(X_i\), yielding
\[
\E[Y_i\mid X_i, Z_g, \bar X_g,  G_i]
= \E[Y_i\mid Z_{G_i}, X_{ij}=1]
= \E[B_{gj}\mid Z_g].
\] Again, these arguments hold because of the multiplication by the
indicator functions outside the expectation. Substituting, we can pull
\(\E[\vb b_{xg}\mid Z_g]\) out, yielding \[
\begin{aligned}
\E[B_{gj}\mid Z_g, \bar X_g]
&= \bar X_{gj}^{-1} \E\left[\sum_i N_g^{-1}\ind\{G_i=g\} x_{xi}
   \E[B_{gj}\mid Z_g] \,\middle|\, Z_g, \bar X_g\right] \\
&= \bar X_{gj}^{-1} \E[B_{gj}\mid Z_g]
    \E\left[\sum_i N_g^{-1}\ind\{G_i=g\} x_{xi} \,\middle|\, Z_g, \bar X_g\right] \\
&= \bar X_{gj}^{-1} \E[B_{gj}\mid Z_g] \bar X_{gj}
= \E[B_{gj}\mid Z_g].
\end{aligned}
\] The result follows by substituting in \(G\) for \(g\).
\end{proof}

\subsection{\texorpdfstring{Proof of Proposition~\ref{prp-subspace} and
Proposition~\ref{prp-msc}}{Proof of Proposition~ and Proposition~}}\label{proof-of-prp-subspace-and-prp-msc}

Throughout this section we drop the subscript \(G\) for clarity. Both
results rely on the following bound.

\begin{lemma}[]\protect\hypertarget{lem-gamma2-lb}{}\label{lem-gamma2-lb}

Let \(\{\eta_j\}_{j=1}^d\in L^2(Z)\). Then under
Assumption~\ref{asm-pos}, for any \(j\) \[
\E[\eta_j(Z)^2] \le \frac{d(d+1)}{\delta(d-1)!} \E[(\eta(Z)^\top \bar X)^2],
\] where \(\delta\) is the lower bound on the density in
Assumption~\ref{asm-pos}.

\end{lemma}

\begin{proof}
First, \[
\E[\eta_j(Z)^2] \le \E\qty[\sum_{j\in\X} \eta_j(Z)^2] =: \norm{\eta}^2,
\] so it suffices to bound \(\norm{\eta}^2\). For this, let \(\nu\) be
the dominating measure in Assumption~\ref{asm-pos}. We have \[
\E[(\eta(Z)^\top \bar X)^2]
=\int_{\Delta^d\times\supp(Z)} (\eta(z)^\top x)^2 f(x, z) \nu(dx, dz).
\]

Since there exists a \(\delta>0\) such that \(f(\bar x, z)>\delta f(z)\)
for all \((\bar x, z)\in\Delta^d\times\supp(Z)\), \[
\begin{aligned}
\E[(\eta(Z)^\top \bar X)^2]
&\ge \int_{\Delta^d\times\supp(Z)} (\eta(z)^\top \bar x)^2 \delta f(z) \nu(d\bar x, dz) \\
&= \delta (d-1)! \int_{\Delta^d\times\supp(Z)} (\eta(z)^\top \bar x)^2 (d-1)!^{-1}f(z) \nu(d\bar x, dz) \\
&= \delta (d-1)! \E[(\eta(Z)^\top W)^2],
\end{aligned}
\] where \(W\) is uniform on \(\Delta^d\), which has volume
\(1/(d-1)!\), independent of \(Z\), and \(\nu\) is the dominating
measure on \(\Delta^d\times\supp(Z)\). Then, since
\(\E[(\eta(Z)^\top W)^2]<\infty\), \[
\begin{aligned}
\E[(\eta(Z)^\top W)^2]
&= \E[\eta(Z)^\top \E[WW^\top\mid Z]\eta(Z)] \\
&= \E[\eta(Z)^\top \E[WW^\top] \eta(Z)] \\
&\ge \lambda_\text{min}(\E[WW^\top])\norm{\eta}^2,
\end{aligned}
\] where \(\lambda_\text{min}\) is the minimum eigenvalue. This
eigenvalue can be computed in closed form: since
\(W\sim\text{Dirichlet}(1,\dots,1)\), by well-known properties of the
Dirichlet distribution, \[
\E[W_i^2] = \frac{2}{d(d+1)} \qand \E[W_iW_j]= \frac{1}{d(d+1)},
\] so \(\E[WW^\top] = \frac{1}{d(d+1)}(I + J)\), where \(J\) is a matrix
of all ones. We claim the eigenvalues of \(I+J\) are \(d+1\) with
multiplicity \(1\) and \(1\) with multiplicity \(d-1\). To see the
latter, \((I+J-1\cdot I)=J\), which has rank \(1\), so we can find
\(d-1\) linearly independent eigenvectors in its null space. To see the
former, we have that \[
2d = \tr(I+J) = \sum_{i=1}^d \lambda_i = (d-1)\cdot 1 + \lambda_d,
\] so \(\lambda_d=d+1\). Thus
\(\lambda_\text{min}(\E[WW^\top])=\frac{1}{d(d+1)}>0\).

Finally, we can substitute and rearrange to find \[
\norm{\eta}^2 \le \frac{d(d+1)}{\delta(d-1)!}\E[(\eta(Z)^\top X)^2].
\]
\end{proof}

\subparagraph{\texorpdfstring{Proof of
Proposition~\ref{prp-msc}}{Proof of Proposition~}}\label{proof-of-prp-msc}

\begin{proof}
We wish to show that
\(\E[\gamma(e_j, Z)^2u(N, \bar X_j)^2]\le C_{\Pr}\norm{\gamma}^2\) for
some constant \(C_{\Pr}\) depending on \(\Pr\). Since
\(E[u(N, \bar X_j)]\) is bounded by Assumption~\ref{asm-bnd},we can
absorb it into \(C_{\Pr}\). It therefore suffices to show that \[
\E[\gamma(e_j, Z)^2] = \E[\eta_j(Z)^2]
\le C_{\Pr}\norm{\gamma}^2  = C_{\Pr}\E[(\eta(Z)^\top \bar X)^2].
\] This is immediate from Lemma~\ref{lem-gamma2-lb} with \[
C_{\Pr}=\frac{C_N^2d(d+1)}{\delta(d-1)!}<\infty.
\]
\end{proof}

\subparagraph{\texorpdfstring{Proof of
Proposition~\ref{prp-subspace}}{Proof of Proposition~}}\label{proof-of-prp-subspace}

\begin{proof}
That \(\Gamma\) is a linear subspace is immediate: for
\(\gamma_1, \gamma_2\in\Gamma\), with corresponding \(\eta_1, \eta_2\),
and \(a,b\in\R\), \[
a\gamma_1 + b\gamma_2
= a\eta_1(Z)^\top \bar X + b\eta_2(Z)^\top \bar X
= (a\eta_1(Z) + b\eta_2(Z))^\top \bar X\in\Gamma.
\] To see closure, take a sequence \(\gamma_n\in\Gamma\) with
\(\gamma_n\to \gamma^*\in L^2(\bar X, Z)\). We can write each
\(\gamma_n\) as \(\gamma_n(\bar x, z)=\eta_{n}(z)^\top \bar x\) where
each \(\eta_{nj}\in L^2(Z)\).

We first claim that there exist \(\eta^*_j\in L^2(Z)\) such that
\(\eta_{nj}\to^{L^2} \eta^*_j\). Since \(L^2(Z)\) is complete and
closed, for the existence of \(\eta^*_j\) it suffices to show that
\(\eta_{nj}\) is Cauchy in \(L^2(Z)\) for each \(j\). Since \(\gamma_n\)
converges, it is Cauchy. Fix an \(\eps>0\); there exists an \(N_\eps\)
such that for all \(m,n>N\) \[
\eps > \E[(\gamma_m(\bar X, Z) - \gamma_n(\bar X, Z))^2]
= \E[((\eta_m(Z)-\eta_n(Z))^\top\bar X)^2];
\] Showing that \(\E[(\eta_{mj}(Z)-\eta_{nj}(Z))^2]\) is bounded by a
constant times \(\eps\) will establish that \(\eta_{nj}\) is Cauchy. But
this follows immediately from Lemma~\ref{lem-gamma2-lb} since all
\(\eta_{mj}(Z)-\eta_{nj}(Z)\in L^2(Z)\).

We now show that \(\gamma_n(\bar x, z)\to^{L^2} \eta^*(z)^\top \bar x\).
By Cauchy-Schwarz and since \(\bar X\) lies in the unit simplex, \[
\begin{aligned}
\E[(\sum_j \eta_{nj}(Z)\bar X_j - \sum_j \eta^*_j(Z)\bar X_j)^2]
&\le \E[(\sum_j (\eta_{nj}(Z)-\eta^*_j(Z))^2)(\sum_j \bar X_j^2)] \\
&\le \sum_j  \E[(\eta_{nj}(Z)-\eta^*_j(Z))^2] \to 0.
\end{aligned}
\] Thus \(\gamma^*(\bar x, z)=\eta^*(z)^\top\bar x\in\Gamma\).
\end{proof}

\subsection{\texorpdfstring{Proof of
Theorem~\ref{thm-sieve}}{Proof of Theorem~}}\label{proof-of-thm-sieve}

\begin{proof}
Let \[
\begin{aligned}
\ell_\gamma(\gamma) := (\bar Y_g - \gamma(\bar X_g, Z_g))^2 \qand
\ell_\alpha(\alpha) := \alpha^2(\bar X_g, Z_g) - 2u(N_g, \bar X_{gj})\alpha(e_j, Z_g)
\end{aligned}
\] be the loss functions, and
\(L_\gamma(\gamma):=\E[\ell_\gamma(\gamma)]\) and
\(L_\alpha(\alpha):=\E[\ell_\alpha(\alpha)]\) be their expectations.

We first handle the unpenalized case, \[
\norm{\hat\gamma_m(0) - \gamma_0} = O_{\Pr}(\rho_m + \sqrt{J_m/m}) \qand
\norm{\hat\alpha_{mj}(0) - \alpha_{0j}} = O_{\Pr}(\rho_m + \sqrt{J_m/m}),
\] which follows from Theorem 3.5 of \citet{chen2007large} once we
verify the following conditions from that work:

\begin{enumerate}
\def\labelenumi{(\arabic{enumi})}
\setcounter{enumi}{8}
\tightlist
\item
  \(\norm{\cdot}\) is equivalent to \(\norm{\cdot}_{2,\nu}\). This
  follows from Assumption~\ref{asm-pos} and Assumption~\ref{asm-sr} (1),
  since they ensure that the joint density satisfies
  \(0<f(\bar x, z)<\infty\) on the (compact) support of those variables.
\item
  \(\Gamma_m\) is identifiable.
\item
  \(\norm{\gamma_0}_\infty\) and \(\norm{\alpha_{0j}}_\infty\) are
  finite. This follows from Assumption~\ref{asm-sr} (3) since the
  \(\bar X\) are bounded.
\item
  For any bounded \(f_1,f_2\in\Gamma^\F\), both
  \(L_\gamma(f_1 + \tau(f_2 - f_1))\) and
  \(L_\alpha(f_1 + \tau(f_2 - f_1))\) are twice continuously
  differentiable w.r.t. \(\tau\in[0, 1]\). Moreover, for any
  \(0<K<\infty\), when \(\norm{f_1}_\infty<K\) and
  \(\norm{f_2}_\infty<K\), these second derivatives are of the same
  order as \(\norm{f_1-f_2}^2\) for all \(\tau\in[0, 1]\).
\item
  For any \(f_1,f_2\in\Gamma_m\), both
  \(\E_m[\ell_\gamma(f_1 + \tau(f_2 - f_1))[\) and
  \(\E_m[\ell_\alpha(f_1 + \tau(f_2 - f_1))]\) are twice continuously
  differentiable w.r.t. \(\tau\in[0, 1]\). Moreover, (i) for
  \(\bar f_{\gamma m} := \arg\min_{f\in\Gamma_m} L_\gamma(f)\) and
  \(\bar f_{\alpha m} := \arg\min_{f\in\Gamma_m} L_\alpha(f)\), \[
  \begin{aligned}
  \sup_{f\in\Gamma_m} \frac{|\partial_\tau \E_m[\ell_\gamma(\bar f_{\gamma m} + \tau f)]|_{\tau=0}|}{\norm{f}}
      &= O_{\Pr}(\sqrt{J_m/m}) \qand \\
  \sup_{f\in\Gamma_m} \frac{|\partial_\tau \E_m[\ell_\alpha(\bar f_{\alpha m} + \tau f)]|_{\tau=0}|}{\norm{f}}
      &= O_{\Pr}(\sqrt{J_m/m}),
  \end{aligned}
  \] and (ii), for any \(0<K<\infty\), there exists a \(c>0\) such that
  \(\norm{f_1}_\infty<K\) and \(\norm{f_2}_\infty<K\), these second
  derivatives upper bounded by \(c \norm{f_1-f_2}^2\) for
  \(\tau\in[0, 1]\) with probability approaching one as \(m\to\infty\).
\item
  For every \(m\), \[
  \sup_{f'\in\Gamma^\F}\inf_{f\in\Gamma_m} \norm{f - f'}_{2,\nu} \le \rho_{m} \qand
  \sup_{f\in\Gamma_m, \norm{f}_{2,\nu}\neq 0} \norm{f}_\infty / \norm{f}_{2,\nu} \le C_A\cdot A_m,
  \] for some universal constant \(C_A\).
\end{enumerate}

As noted above, (9) and (11) are immediate. For (10), let
\(f\in\Gamma_m\), so we have a \(\theta\in\R^{dJ_m}\) with \[
f(\bar x, z)=(\Phi_m(z)\otimes \bar x)^\top\theta
= \sum_{j=1}^d \bar x_j \Phi_m(z)^\top \theta_j,
\] where \(\theta_j\) is the corresponding subvector of \(\theta\). For
a given \(z\), the above representation is a dot product between a
vector \(\bar x\) and a vector with elements
\(\Phi_m(z)^\top \theta_j\). The set of vectors \(\bar x\) which can
make this dot product zero is a hyperplane of dimension \(d-1\), which
is measure zero in \(\Delta^d\). Thus if \(\norm{f}_{2,\nu}=0\), then we
must have each \(\Phi_m(z)^\top \theta_j=0\). But then
Assumption~\ref{asm-sr} (4) in turn implies that each \(\theta_j=0\),
and so \(f=0\).

For (12) we compute, for any \(f,f'\in L^2(\bar X, Z)\) (so that we can
differentiate inside the expectation), \[
\begin{aligned}
\partial_\tau L_\gamma(f + \tau f')
&= \E[-2(\bar Y_g - (f + \tau f'))f'] \qand
\partial^2_\tau L_\gamma(f + \tau f') = \E[2f'^2], \\
\partial_\tau L_\alpha(f + \tau f')
&= \E[2(f+\tau f')f' - 2u f'(e_j, Z_g)] \qand
\partial^2_\tau L_\alpha(f + \tau f') = \E[2f'^2].
\end{aligned}
\] where we have suppressed the dependence of the functions on the data
for clarity.

Letting \(f=f_1\) and \(f'=f_2-f_1\) for any bounded
\(f_1,f_2\in \Gamma^\F\subseteq L^2(\bar X, Z)\), (12) is clearly
satisfied. The same derivation yields the same derivatives for
\(\ell_\gamma\) and \(\ell_\alpha\), but without the outer expectation,
which make clear that for
\(f_1,f_2\in \Gamma_m\subseteq L^2(\bar X, Z)\), the first part of (13)
as well as (13)(ii) are satisfied.

For (13)(i), first note that we must have
\(\partial_\tau L_\gamma(\bar f_{\gamma m} + \tau f)=0\), since
\(\bar f_{\gamma m}\) minimizes \(L_\gamma\); otherwise, because
\(L_\gamma\) is continuously differentiable, we could reduce
\(L_\gamma\) by moving in the direction \(-f\). Thus we can equivalently
show that the condition holds with \(\E_m\) replaced by \(\E_m - \E\).
Combining this observation with Remark A.1 of \citet{huang2001concave},
it suffices to show \(|S(\theta)|=O_{\Pr}(\sqrt{J_m/m})\), where \(S\)
is the empirical process \[
S_i(\theta) := \partial_{\theta_i} (\E_m - \E)[\ell_\gamma(\phi(\bar X, Z)^\top\theta)]
\] for each \(i\in\X\), where \(\{\phi_k\}_{k=1}^{dJ_m}\) are an
orthonormal basis of \(\Gamma_m\). Since
\(|S(\theta)|^2=\sum_{k=1}^{dJ_m} |S_k(\theta)|^2\), it suffices to show
that each \(S_i(\theta)^2=O_{\Pr}(m^{-1})\), i.e., that \[
\frac{1}{m} (m^{-1/2}\sum_{g=1}^m (\partial_{\theta_i} \ell_\gamma - \E[\partial_{\theta_i} \ell_\gamma]))^2
= O_{\Pr}(m^{-1}).
\] This is immediate from the central limit theorem if
\(\E[\partial_{\theta_i} \ell_\gamma]^2\) is finite. For
\(\ell_\gamma\), we have \[
\partial_{\theta_i} \ell_\gamma
= -2(\bar Y_g - \phi(\bar X_g, Z_g)^\top\theta)\phi_i(\bar X_g, Z_g);
\] \(\bar Y_g\) has finite variance by Assumption~\ref{asm-sr} (2) since
\(\gamma_0\) is bounded,
\(\phi(\bar X_g, Z_g)^\top\theta=\bar f_{\gamma m}\), which is bounded
by (11) and (12) (see \citet{huang2001concave}, Theorem A.1), and
\(\phi_i(\bar X_g, Z_g)\in \Gamma_m\) and so is bounded. Thus
\(\E[\partial_{\theta_i} \ell_\gamma]^2\) is finite. For
\(\ell_\alpha\), we have \[
\partial_{\theta_i} \ell_\alpha
= 2\phi(\bar X_g, Z_g)^\top\theta\cdot \phi_i(\bar X_g, Z_g) - 2u(N_g, \bar X_{gj})\phi_i(Z_g, e_i);
\] \(\phi(\bar X_g, Z_g)^\top\theta=\bar f_{\alpha m}\) is bounded by
(11) and (12), \(\phi_i\in \Gamma_m\) and so is bounded, and
\(u(N_g, \bar X_{gj})\) is bounded by Assumption~\ref{asm-bnd}. Thus
\(\E[\partial_{\theta_i} \ell_\alpha]^2\) is finite as well, and so
(13)(i) holds.

Finally, for (14), take \(f'\in\Gamma^\F\), so there exist \(f'_j\in\F\)
with \(f'(\bar x, z) = \sum_j \bar x_j f'_j(z)\). Then, representing
\(f\in\Gamma_m\) similarly, with \(f_j\in\span\Phi_m\), \[
\begin{aligned}
\inf_{f\in\Gamma_m} \norm{f - f'}_{2,\nu}
&= \inf_{\{f_j\}\in\span\Phi_m} \norm{\sum_j \bar X_j (f_j - f_j')}_{2,\nu} \\
&\le \sum_{j=1}^d \inf_{f_j\in\span\Phi_m} \norm{\bar X_j (f_j - f_j')}_{2,\nu} \\
&= \sum_{j=1}^d \inf_{f_j\in\span\Phi_m} \norm{\bar X_j}_{2,\nu}\norm{(f_j - f_j')}_{2,\nu} \\
&= \inf_{f_j\in\span\Phi_m} \norm{(f_j - f_j')}_{2,\nu} = \rho_m,
\end{aligned}
\] since \(\bar X\) and \(Z\) are independent in \(\nu\) and
\(\norm{\bar X_j}_{2,\nu}=d^{-1}\). Since this holds for every
\(f'\in\Gamma^\F\), the first part of (14) holds. For the second part,
let \(f\in\Gamma_m\) with \(\norm{f}_{2,\nu}\neq 0\), so there exist
\(f_j\in\span\Phi_m\) with \(f(\bar x, z)=\sum_j \bar x_j f_j(z)\). Then
since \(\bar x\in\Delta^d\), \[
\norm{f}_\infty = \max_j \norm{f_j}_\infty
\le A_m \max_j \norm{f_j}_{2,\nu} \le C_A A_m \norm{f}_{2,\nu},
\] where the second equality applies the definition of \(A_m\) and the
final equality follows from Lemma~\ref{lem-gamma2-lb} since
\(\norm{\cdot}\) and \(\norm{\cdot}_{2,\nu}\) are equivalent. To apply
the definition of \(A_m\), we must have at least one
\(\norm{f_j}_{2,\nu}\neq 0\). This is true because \[
\begin{aligned}
\norm{f}_{2,\nu}^2
&= E_\nu[f(\bar X, Z)^\top \E_\nu[\bar X\bar X^\top] f(\bar X, Z)] \\
&\le \lambda_{\max(\E_\nu[\bar X\bar X^\top])} E_\nu[f(\bar X, Z)^\top f(\bar X, Z)]
= \frac{1}{d}\sum_{j=1}^d \norm{f_j}_{2,\nu}^2,
\end{aligned}
\] so if the left-hand side is nonzero, then at least one
\(\norm{f_j}_{2,\nu}^2\) must be nonzero as well. Thus the second part
of (14) holds as well, and so the theorem holds in the unpenalized case.

To handle penalization, we will show that the condition on \(\lambda_m\)
implies that penalization has no asymptotic effect. Specifically, we
will show that \(\E_m[(\hat\gamma(0) - \hat\gamma(\lambda_m))^2]\) and
\(\E_m[(\hat\alpha_{0j}(0) - \hat\alpha_{0j}(\lambda_m))^2]\) are both
\(O_{\Pr}(J_m/m)\), where \(\hat{\vb w}_{mj}\) is the vector of
evaluated Riesz weights \(\hat\alpha_{mj}(\bar X_{gj}, Z_g)\) for
\(g=1,\dots, m\). By the triangle inequality and the proof of Theorem
3.5 of \citet{chen2007large} (i.e., \citet{huang2001concave}, Theorem
A.2), this condition sufficient is to establish the main result for the
penalized estimators, whose estimation error is \(O(\sqrt{J_m/m})\). We
can write these differences as \[
\begin{aligned}
\E_m[(\hat\gamma(0) - \hat\gamma(\lambda_m))^2]
&= m^{-1}\norm{U_mD_m^2D_m^{-2}U_m^\top\vb y - U_mD_m^2(D_m^2+\lambda_m I)^{-1}U_m^\top\vb y_m}^2 \\
&= m^{-1}\norm{U_mD_m^2(D_m^{-2}-(D_m^2+\lambda_m I)^{-1})U_m^\top\vb y_m}^2 \\
&\le \norm{U_m}_\op^2 \norm{D_m^2(D_m^{-2}-(D_m^2+\lambda_m I)^{-1})}_\op^2 \norm{U_m^\top}_\op^2 \norm{\vb y_m}^2/m \\
&= \norm{D_m^2(D_m^{-2}-(D_m^2+\lambda_m I)^{-1})}_\op^2 \norm{\vb y_m}^2/m \\
&= (\max_k (1 - \frac{d_{mk}^2}{d_{mk}^2 + \lambda_m}))^2 \norm{\vb y_m}^2/m \\
&= (1 - \frac{d_{mJ_m}^2}{d_{mJ_m}^2 + \lambda_m})^2 \norm{\vb y_m}^2/m
\end{aligned}
\] for \(\gamma_0\), where \(U_mD_mV_m^\top\) is the singular value
decomposition of the design matrix \(\XZ_m\) with rows
\(X_{gj}\otimes \Phi(Z_g)\), and \[
\begin{aligned}
\E_m[&(\hat\alpha_{0j}(0) - \hat\alpha_{0j}(\lambda_m))^2] \\
&= m^{-1}\norm{U_mD_mD_m^{-2}V_m^\top (E'\XZjmp)^\top\ind - U_mD_m(D_m^2+\lambda_m I)^{-1}V_m^\top (E'\XZjmp) \ind}^2 \\
&= m^{-1}\norm{U_mD_m(D_m^{-2}-(D_m^2+\lambda_m I)^{-1})V_m^\top V'_mD'_m{U'_m}^\top E'^\top\ind}^2 \\
&\le \norm{D_mD'_m(D_m^{-2}-(D_m^2+\lambda_m I)^{-1})}_\op^2 \norm{E'^\top\ind}^2/m \\
&\le C'\cdot C_N\cdot (1 - \frac{d_{mJ_m}^2}{d_{mJ_m}^2 + \lambda_m})^2
\end{aligned}
\] for \(\alpha_{0j}\) by the results below, where \(\XZjmp\) is the
modified design matrix with each row \(e_j\otimes \Phi(Z_g)\), with
singular value decomposition \(U'_mD'_m{V'_m}^\top\), and \(E'\) is the
diagonal matrix with entries \(u(N_g,\bar X_{gj})\), so that \(\XZjm\)
below corresponds with \(E'\XZjmp\). We assume the singular values are
ordered \(d_{m1}\ge d_{m2}\ge \dots \ge d_{mJ_m}\). The fact that
\(\norm{E'^\top\ind}^2/m\le C_N\) follows from Assumption~\ref{asm-bnd}.
That we can bound the term with \(D_mD'_m\) as if it were \(D_mD_m\)
follows from the fact that the singular values of a tensor product are
the products of the singular values of each component matrix; the matrix
with \(\bar X\) has singular values bounded below by
Assumption~\ref{asm-pos}, and the singular values of \(\Phi_m(Z)\) are
shared with \(D_m\) and \(D_m'\). So the smallest positive value in
\(D'_m\) is a uniform constant away from the smallest positive value in
\(D_m\).

Now, under Assumption~\ref{asm-sr} (2),
\(\norm{\bar{\vb y}_m}^2/m\cvp\E[\bar Y]\). Thus for both \(\hat\gamma\)
and \(\hat\alpha\) we must show that \[
1 - \frac{d_{mJ_m}^2}{d_{mJ_m}^2 + \lambda_m} = O_{\Pr}(\sqrt{\frac{J_m}{m}}).
\] Since the eigenvalues of \(\Phi_m(\vb Z)^\top\Phi_m(\vb Z)\) are
assumed to be uniformly bounded below, and the eigenvalues of
\(\bar X^\top\bar X\) are bounded below by Assumption~\ref{asm-pos} (see
the above proofs), the eigenvalues of \(\XZ_m^\top \XZ_m\), i.e., any
\(d_{mk}^2\), are uniformly bounded below as well. We then have that \[
\sqrt{\frac{m}{J_m}}(1 - \frac{d_{mJ_m}^2}{d_{mJ_m}^2 + \lambda_m})
= \sqrt{\frac{m}{J_m}} - \frac{\sqrt{m}}{\sqrt{J_m}(1+\lambda_m/d_{mJ_m}^2)},
\] which, since \(d_{mJ_m}^2\) is uniformly bounded below, is bounded if
\(\lambda_m=O(\sqrt{J_m/m})\), which it does by assumption. Therefore
penalization has no asymptotic effect.
\end{proof}

\subsection{\texorpdfstring{Proof of
Theorem~\ref{thm-dml}}{Proof of Theorem~}}\label{proof-of-thm-dml}

\begin{proof}
The result is a direct application of Theorem 1 of
\citet{chen2022debiased}, which requires four conditions: a consistency
rate, Neyman orthogonality, smoothness, and stochastic equicontinuity.
Consistency is by assumption, and Neyman orthogonality is a property of
the score that defines \(\hat\beta\). For smoothness, we must show that
\[
\partial^2_\tau\E[\psi_j(\gamma_0+\tau(\gamma-\gamma_0), \alpha_0+\tau(\alpha-\alpha_0))]\rvert_{\tau=0}
= O(\norm{\gamma-\gamma_0}^2 + \norm{\alpha-\alpha_0}^2).
\] Let \(\tilde\gamma\) denote evaluating \(\gamma\) with \(\bar X_j\)
fixed to \(e_j\), so that \[
\psi_j(\gamma,\alpha) = U\tilde\gamma - \alpha(Y - \gamma)
\] can be written more cleanly. We then have \[
\begin{aligned}
\partial_\tau&\{
U(\tilde\gamma_0+\tau(\tilde\gamma-\tilde\gamma_0))
+(\alpha_0+\tau(\alpha-\alpha_0))(\bar Y - (\gamma_0+\tau(\gamma-\gamma_0))
\} \\
&= U(\tilde\gamma - \tilde\gamma_0) +
    (\alpha-\alpha_0)(\bar Y - \gamma_0 - \tau(\gamma-\gamma_0)) -
    (\alpha_0 + \tau(\alpha-\alpha_0))(\gamma - \gamma_0)
\end{aligned}
\] and \[
\begin{aligned}
\partial^2_\tau&\{
U(\tilde\gamma_0+\tau(\tilde\gamma-\tilde\gamma_0))
+(\alpha_0+\tau(\alpha-\alpha_0))(\bar Y - (\gamma_0+\tau(\gamma-\gamma_0))
\} \\
&= -2\tau(\alpha-\alpha_0)(\gamma - \gamma_0),
\end{aligned}
\] which is zero when evaluated at \(\tau=0\). So the smoothness
condition follows.

\citet{chen2022debiased} establish that stochastic equicontinuity
follows from certain continuity conditions, which they establish for DML
estimators based on Riesz representers like ours, and if the following
algorithmic stability conditions are satisfied:
\begin{equation}\protect\phantomsection\label{eq-stable}{
\begin{aligned}
\max_{i\le m} \E[\sup_{\bar x, z} (\hat\gamma_m(\bar x, z)
    - \hat\gamma^{(-i)}_m(\bar x, z))^{2r}]^{1/2r} &= o(m^{-1/2}) \qand \\
\max_{i\le m} \E[\sup_{\bar x, z} (\hat\alpha_{mj}(\bar x, z)
    - \hat\alpha^{(-i)}_{mj}(\bar x, z))^{2r}]^{1/2r} &= o(m^{-1/2}),
\end{aligned}
}\end{equation} where the \((-i)\) superscript denotes the estimator
computed without observation \(i\).

First, we will express \(\hat\alpha\) as numerically equivalent to a
certain ridge regression on the same data as \(\hat\gamma\) but with a
different outcome, so that establishing both stability conditions
follows from establishing the stability of ridge regression, plus any
conditions on the outcome variable. For this first step, from
Appendix~\ref{sec-app-riesz}, the closed-form solution for
\(\hat\alpha\) is given by parameter vector \[
\theta^* = (\XZ_m^\top \XZ_m + \lambda_m I)^{-1} \XZjm^\top \ind.
\] This is clearly equivalent to ridge regression with design matrix
\(\XZ_m\) if we can find an outcome vector \(\tilde{\vb y}\) such that
\(\XZ_m^\top \tilde{\vb y} = \XZjm^\top \ind\). This is possible since
\(\XZ_m^\top\) has rank \(J_m\), while \(\tilde{\vb y}\) has \(m>J_m\)
entries. We can in fact bound
\(\norm{\tilde{\vb y}}\le \norm{\XZ_m^+}_\op m=O(m)\), since
\(\norm{\XZ_m^+}_\op\) is upper bounded by the assumption that the
eigenvalues of \(\Phi_m(Z)^\top\Phi_m(Z)\) are uniformly bounded away
from zero; see the discussion in the proof of Theorem~\ref{thm-sieve}.
This implies that we can take \(\tilde{\vb y}\) with entries bounded
almost surely.

It remains to show that ridge regression on
\(\Phi_m(Z_g)\otimes\bar X_g\) satisfies Eq.~\ref{eq-stable}. For this,
we will apply Lemma 1 of \citet{celisse2016stability}. Let \(\theta\)
and \(\theta^{(-i)}\) be the parameter estimates for ridge regression on
\(\Phi_m(Z_g)\otimes\bar X_g\) fitted on the full data and with
observation \(i\) removed, respectively, and let \(Y\) be a generic
outcome variable (here, either \(\bar Y\) or \(\tilde Y\)). Lemma 1 of
\citet{celisse2016stability} shows that for any \(0<\chi<1\) \[
\norm{\theta-\theta^{(-i)}}
\le \frac{O(\sqrt{J_m})}{m^2\lambda_m}(|Y_i| + \frac{O(J_m)+m\lambda_m}{m\lambda_m(\chi-1)}
    (\frac{1}{m-1}\sum_{l\neq i} |Y_l|)).
\] For readers cross-referencing the lemma, \citet{celisse2016stability}
parametrize \(\lambda\) differently; their \(\lambda\) corresponds to
our \(m\lambda_m\). Their \(B_X\) is a bound on the 2-norm of the
covariates, which here is just \(O(\sqrt{J_m})\) since each basis
function as well as \(\bar X\) is bounded. We can bound the inside of
the expectation in Eq.~\ref{eq-stable} by Cauchy-Schwarz and the fact
that \(\norm{\Phi_m(z)\otimes\bar x}=O(\sqrt{J_m})\): \[
\sup_{\bar x, z} (\hat\gamma_m(\bar x, z) - \hat\gamma^{(-i)}_m(\bar x, z))^{2r}
= \sup_{\bar x, z} ((\theta -\theta^{(-i)})^\top (\Phi_m(z)\otimes\bar x) )^{2r}
\le \sup_{\bar x, z} (\norm{\theta -\theta^{(-i)}}O(\sqrt{J_m}))^{2r}
\] the same holds for \(\hat\alpha_{mj}\). Taking the expectation and
simplifying the same way as Eq. 7 in \citet{celisse2016stability}, we
have \[
\E[\sup_{\bar x, z} (\hat\gamma_m(\bar x, z) - \hat\gamma^{(-i)}_m(\bar x, z))^{2r}]^{1/2r}
\le \norm{Y}_{2r} \frac{O(J_m)}{m^2\lambda_m}(1+\frac{O(J_m)+m\lambda_m}{m\lambda_m(\chi-1)})
\] Since \(\norm{Y}_{2r}\) is finite by assumption, and
\(\lambda_m\asymp \sqrt{J_m/m}\), we have \[
\begin{aligned}
\E[\sup_{\bar x, z} (\hat\gamma_m(\bar x, z) - \hat\gamma^{(-i)}_m(\bar x, z))^{2r}]^{1/2r}
&= O(\frac{\sqrt{mJ_m}}{m^2})O(\frac{\sqrt{mJ_m}(1 + \sqrt{J_m/m})}{\sqrt{mJ_m}}) \\
&= O(\frac{\sqrt{J_m/m}}{m}) = o(m^{-1/2}).
\end{aligned}
\] Since this does not depend on the left-out observation \(i\), the
proof is complete.
\end{proof}

\subsection{\texorpdfstring{Proof of
Proposition~\ref{prp-var-id}}{Proof of Proposition~}}\label{proof-of-prp-var-id}

\begin{proof}
We first show that
\(\kappa_0(x, z)= \E[(\eps_G^\top \bar X_G)^2\mid \bar X_G=x,Z_G=z]\).
By Assumption~\ref{asm-car}, \[
\begin{aligned}
\E[(\bar Y_G - \E[\bar Y_G \mid \bar X_G,Z_G])^2\mid \bar X_G,Z_G]
&= \E[(B_G^\top\bar X_G - \eta(Z_G)^\top\bar X_G)^2\mid \bar X_G,Z_G] \\
&= \E[((B_G-\eta(Z_G))^\top\bar X_G)^2\mid \bar X_G,Z_G] \\
&= \E[(\eps_G^\top\bar X_G)^2\mid \bar X_G,Z_G].
\end{aligned}
\]

Second, for any \(a,b\), by Assumption~\ref{asm-car2} \[
\begin{aligned}
\kappa(a e_j + b e_k, z)
&= \E[\bar X_G^\top \eps_G\eps_G^\top \bar X_G\mid \bar X_G=ae_j+be_k, Z_g=z] \\
&= (ae_j+be_k)^\top \E[\eps_G\eps_G^\top\mid \bar X_G=ae_j+be_k, Z_g=z] (ae_j+be_k) \\
&= (ae_j+be_k)^\top \E[\eps_G\eps_G^\top\mid Z_g=z] (ae_j+be_k) \\
&= (ae_j+be_k)^\top \Sigma(z) (ae_j+be_k) \\
&= a^2\Sigma_{jj}(z) + b^2\Sigma_{kk}(z) + 2ab\Sigma_{jk}(z).
\end{aligned}
\] Thus \[
\begin{aligned}
2\kappa(\thalf e_j + \thalf e_k, z) &-
    \tquart\kappa(e_j, z) - \tquart\kappa(e_k, z) \\
&= 2((\tquart\Sigma_{jj}(z) + \tquart\Sigma_{kk}(z) + 2\cdot\tquart\Sigma_{jk}(z))
     -\tquart\Sigma_{jj}(z) - \tquart\Sigma_{kk}(z)) \\
&= \Sigma_{jk}(z).
\end{aligned}
\]
\end{proof}

\subsection{\texorpdfstring{Proof of
Theorem~\ref{thm-local-ci}}{Proof of Theorem~}}\label{proof-of-thm-local-ci}

\begin{proof}
\leavevmode

Fix \(g\) and let \(\norm{\,\cdot\,}_{\hat\Pi\hat\Sigma}\) denote the
(random) seminorm \(x\mapsto x^\top(\hat\Pi_g\hat\Sigma(Z_g))^+x\).
Since \(H'(\bar x, \bar y)\) is convex, we have for any \(c\in\R^d\) and
\(b\in H'(\bar x, \bar y)\) that \[
\norm{c^* - b}_{\hat\Pi\hat\Sigma}\le \norm{c - b}_{\hat\Pi\hat\Sigma},
\] where \(c^*\) is the oblique projection of \(c\) onto
\(H'(\bar x, \bar y)\) along \(\hat\Sigma(Z_g)\). Since
\(B_g\in H'(\bar x, \bar y)\) by construction, \[
\begin{aligned}
1 - \Pr(B_g\in {R'_g}^\alpha)
&= \Pr(\norm{B_g - \hat B'_g}_{\hat\Pi\hat\Sigma} > \frac{d-1}{\alpha}) \\
&\le \frac{\alpha}{d-1} \E[\norm{B_g - \hat B'_g}_{\hat\Pi\hat\Sigma}] \\
&\le \frac{\alpha}{d-1} \E[\norm{B_g - \hat B_g}_{\hat\Pi\hat\Sigma}] \\
&= \frac{\alpha}{d-1} \E[\tr( (\hat\Pi_g\hat\Sigma(Z_g))^+\, (B_g - \hat B_g)(B_g - \hat B_g)^\top )] \\
&= \frac{\alpha}{d-1} \E[\tr( (\hat\Pi_g\hat\Sigma(Z_g))^+\,
    \hat\Pi_g\E[(B_g - \hat\eta(Z_g))(B_g - \hat\eta(Z_g))^\top\mid Z_g])] \\
&= \frac{\alpha}{d-1} \E[\tr( (\hat\Pi_g\hat\Sigma(Z_g))^+\, (\hat\Pi_g\Sigma(Z_g)) )].
\end{aligned}
\] Then since \(\hat\Sigma(Z_g)\cvp \Sigma(Z_g)\) by assumption, by the
continuous mapping theorem \[
\tr((\hat\Pi_g\hat\Sigma(Z_g))^+\, (\hat\Pi_g\Sigma(Z_g)) )
\cvp \tr((\Pi_g\hat\Sigma(Z_g))^+\, (\Pi_g\Sigma(Z_g)) )
= d - 1,
\] where \(\Pi_g\) is the limiting projection matrix, which exists again
by the continuous mapping theorem. That the final trace is \(d-1\)
follows because \(\Pi_g\Sigma(Z_g)\) has rank \(d-1\). Substituting this
into the above expression and cancelling the \(d-1\), we have the
desired result.
\end{proof}

\section{Validation study details}\label{sec-valid-detail}

\subsection{Simulation study details}\label{simulation-study-details}

\subsubsection{Data generating process}\label{data-generating-process}

We generate data from the following truncated Normal ecological model,
which nests the model of \citet{king1997solution}. The model is
implemented in the \texttt{ei\_synthetic} function of our \texttt{seine}
package. \[
\begin{aligned}
     \mqty{\bar X_g\\ Z_g} &\iid
         \mathcal{N}_{\Delta^d \times \R^p}\left(
         \mqty{\mu_x\\ 0},
         \mqty{\Sigma_x & \Gamma \\ \Gamma & T}\right) \\
     \eta &= Z_g^\top \Lambda + \mu_b \\
     B_g &\iid \mathcal{N}_{[0, 1]^d}(\eta, \Sigma_b) \\
     \bar Y_g &= B_g^\top \bar X_g,
\end{aligned}
\] where \(\mu_x\) and \(\Sigma_x\) are the mean and covariance of the
Normal approximation to a Dirichlet distribution, and \(\Gamma\), \(T\),
and \(\Lambda\) are matrices sampled to have certain properties, as
described below. The subscripts on \(\mathcal{N}\) indicate truncation;
i.e., both the predictors \(\bar X\) and the local parameters \(B\) are
truncated to the \(d\)-dimensional hypercube.

The Dirichlet distribution which \(\mu_x\) and \(\Sigma_x\) approximate
has concentration parameter \(\alpha_j=j\). This creates a range of
sizes of predictor groups.

The matrix \(T\) is a symmetric Toeplitz matrix with diagonals
\((0.25\exp(-(k-1)/2))_{k=1}^p\). For example, when \(p=3\), the main
diagonal has value \(0.25\), the next diagonal has value \(0.15163\),
and the final diagonal (which is just one entry in each corner) has
value \(0.09197\). This decreasing sequence is sufficient for a positive
definite \(T\).

The matrices \(\Gamma\) and \(\Lambda\) are initially filled with
independent samples from a standard Normal distribution. \(\Gamma\) is
then numerically projected so that its rows sum to zero, preserving the
sum-to-1 requirement on \(\bar X\), and so that its columns are scaled
to produce the correct \(R^2_{\bar X\sim Z}\). The matrix \(\Lambda\) is
likewise scaled to produce the correct \(R^2_{B\sim Z}\). Due to the
truncation in the sampling of \(\bar X\) and \(B\), the in-sample
\(R^2\) values may be slightly smaller than the ones declared as
simulation parameters.

\subsubsection{Simulation studies}\label{simulation-studies}

\paragraph{Study 1}\label{study-1}

With the setup described above, we generate 1000 Monte Carlo replicates
of datasets with the following parameters:

\begin{itemize}
\tightlist
\item
  \(m=500\), \(d=2\), and \(p=3\)
\item
  \(\mu_b=(0.3, 0.7)\),
\item
  \(\Sigma_b = 0.02(I + \ind\ind^\top)\)
\item
  \(R^2_{\bar X\sim Z}=R^2_{B\sim Z}=0.5\).
\end{itemize}

Our proposed DML estimator uses the \(p = 3\) covariates, entered
linearly. We do not constrain out estimates to fall in the {[}0, 1{]}
square. We used the latest CRAN versions of King's ei (from the
\texttt{ei} package) and Rosen et al.'s \texttt{ei.MD.bayes} (from the
\texttt{eiPack} package). Both functions accept a set of covariates as
an optional argument.

\paragraph{Study 2}\label{study-2}

In addition to the setup described above and in the methods section, we
use:

\begin{itemize}
\tightlist
\item
  \(m\in\{50, 100, 500, 1\,000, 10\,000\}\)
\item
  \(d\in\{2,5,10\}\)
\item
  \(p\in\{1,3,10\}\)
\item
  \(\mu_b\) linearly spaced from \(0.3\) to \(0.7\),
\item
  \(\Sigma_b = 0.005(I + \ind\ind^\top)\)
\item
  \(R^2_{\bar X\sim Z}\in\{0, 0.2, 0.5\}\)
\item
  \(R^2_{B\sim Z}=0.2\)
\end{itemize}

resulting in \(5\times 3\times 3\times 3 = 135\) settings. We generate
estimates with our proposed DML estimates as in Study 1.

\paragraph{Study 3}\label{study-3}

We use the same data generating process as Study 2. Because local
estimates take longer to estimate and each data replicate already
contains \(m\times d\) estimates, we limit our number of replicates to
300. The error and coverage statistics are computed by averaging the
error coverage of local estimates within each simulation replicate and
then across replicates. We specify that local estimates must sum to 1
within each precincts, and assume unimodal residual distribution to
estimate confidence intervals.

\subsection{Voter file validation
details}\label{voter-file-validation-details}

We constructed our dataset with a voter file distributed from the
Florida Department of Elections from January 2024. The file contains
county and precinct identifiers, and an individual's party and racial
affiliation. The major options for party registration are Republican,
Democrat, and Non-Party affiliation. We collapsed the eight available
racial categories into five: White (non-Hispanic), Black (non-Hispanic),
Hispanic (any-part), Asian, and all others (including unknown and
multi-racial). We then linked the precinct identifiers to its geographic
coordinates \citep{vest} and using that as a crosswalk, collected the
following covariates for each precinct:

\begin{itemize}
\tightlist
\item
  population density
\item
  proportion of 2022 votes for U.S. Senate won by Marco Rubio (R)
\item
  proportion of 2022 votes for Governor won by Ron DeSantis
\item
  difference between these two proportions (measuring ticket
  splitting)\footnote{Of course, without penalization this coefficient
    would be aliased, but with penalization, it may aid prediction.}
\item
  proportion of households in poverty
\item
  log median income
\item
  proportion of adults with a college degree or higher
\item
  proportion of White adults with a college degree or higher
\item
  proportion of White adults owning their homes
\item
  proportion of non-White adults owning their homes
\item
  proportion of Hispanics who are of Cuban origin, are of Dominican
  origin, are of Mexican origin, are of Puerto Rican origin
\item
  proportion of adults aged between 18 and 35
\item
  proportion of adults aged above 65
\end{itemize}

The density and election covariates were taken from \citet{vest}. All
other covariates were taken from the U.S. Census Bureau's American
Community Survey estimates at the tract-level (with estimates from the
nearest or encompassing tract imputed onto the precinct). We dropped
precincts which we could not match to an election result or Census
tract.

\subsection{Pollution application}\label{sec-pollution}

Data on race and income by ZCTA was obtained from the 2016 American
Community Survey 5-year estimates. All covariates except for the
latitude and longitude are obtained from the author's analysis. We limit
our analysis to the 2016 data for computational simplicity.

In the contour plots, benchmarking was run for each racial group within
the \$0-20k category,and then the maximum value of the sensitivity
parameters across the two racial groups was taken as the benchmarked
value for the difference.

\end{document}